\documentclass{article}

\usepackage[T1]{fontenc}
\usepackage[dvipsnames]{xcolor}
\usepackage{microtype}
\usepackage{graphicx}
\usepackage{subfigure}
\usepackage{booktabs} 
\usepackage{soul}
\usepackage{amsmath}
\usepackage{amssymb}
\usepackage{mathtools}
\usepackage{amsthm}
\usepackage{physics,amsfonts,bm,braket}
\usepackage{enumerate}

\usepackage{url}
\makeatletter
\g@addto@macro{\UrlBreaks}{\UrlOrds}
\makeatother

\usepackage{hyperref}

\usepackage[accepted]{arxiv2024}

\usepackage[capitalize,noabbrev]{cleveref}

\theoremstyle{plain}
\newtheorem{theorem}{Theorem}[section]
\newtheorem{proposition}[theorem]{Proposition}

\theoremstyle{definition}

\theoremstyle{remark}

\usepackage[textsize=tiny]{todonotes}

\DeclareMathOperator{\erfc}{erfc}

\DeclareMathOperator{\sgn}{sgn}

\DeclareMathOperator*{\extr}{extr}

\newcommand*{\diff}{\mathop{}\!\mathrm{d}}
\newcommand*{\Diff}{\mathop{}\!\mathcal{D}}
\newcommand{\mean}[1]{\left\langle #1 \right\rangle}
\newcommand{\E}{\mathbb{E}}
\newcommand{\normLtwo}[1]{\lVert \vb{#1} \rVert^2_2 }
\newcommand{\normLtwonosq}[1]{\lVert \vb{#1} \rVert_2 }

\newcommand{\eg}{{\em e.g.}}
\newcommand{\ie}{{\em i.e.}}

\arxivtitlerunning{Principled under/oversampling of data for optimal classification}

\begin{document}

\twocolumn[
\arxivtitle{Restoring balance: principled under/oversampling of data for optimal classification}

\arxivsetsymbol{equal}{*}

\begin{arxivauthorlist}
\arxivauthor{Emanuele Loffredo}{equal,ens}
\arxivauthor{Mauro Pastore}{equal,ens}
\arxivauthor{Simona Cocco}{ens}
\arxivauthor{R\'{e}mi Monasson}{ens}
\end{arxivauthorlist}

\arxivaffiliation{ens}{Laboratoire de physique de l'\'{E}cole normale sup\'{e}rieure, CNRS-UMR8023,
PSL University, Sorbonne University, Universit\'{e} Paris-Cit\'{e}
24 rue Lhomond, 75005 Paris, France}

\arxivcorrespondingauthor{Emanuele Loffredo}{emanuele.loffredo@phys.ens.fr}
\arxivcorrespondingauthor{Mauro Pastore}{mauro.pastore@phys.ens.fr}

\arxivkeywords{Imbalanced learning}

\vskip 0.3in
]

\printAffiliationsAndNotice{\arxivEqualContribution} 

\begin{abstract}
    Class imbalance in real-world data poses a common bottleneck for machine learning tasks, since achieving good generalization on under-represented examples is often challenging. Mitigation strategies, such as under or oversampling the data depending on their abundances, are routinely proposed and tested empirically, but how they should adapt to the data statistics remains poorly understood.
    In this work, we determine exact analytical expressions of the generalization curves in the high-dimensional regime for linear classifiers (Support Vector Machines). We also provide a sharp prediction of the effects of under/oversampling strategies depending on class imbalance, first and second moments of the data, and the metrics of performance considered. We show that mixed strategies involving under and oversampling of data lead to performance improvement. Through numerical experiments, we show the relevance of our theoretical predictions on real datasets,  on deeper architectures and with sampling strategies based on unsupervised probabilistic models.
\end{abstract}

\section{Introduction\label{sec:intro}}
 
Many real-world classification tasks, {\em e.g.} in automated medical diagnostics~\citep{krawczyk2016cancer,fotouhi2019cancer}, molecular biology~\citep{wang2006ncRNA,yang2012gene,cheng2015proteinRNA,song2021proteinFunc,ansari2023peptide}, text classification~\citep{liu2009text,liu2017twitter}, ... are plagued by the so-called curse of class imbalance~\citep{kubat1997curse,lemaitre2017curse}: one or more classes in the training data are significantly under-represented, yet correct identification of these rare examples is crucial for performance~\citep{weiss2004rarity}. For such imbalanced datasets, machine learning methods struggle to achieve good classification performances when tested fairly~\citep{he2013book}. 

Due to its wide-spread and intrinsic nature, the issue of learning with class imbalance has long been studied in the computer science literature.
Systematic empirical studies, assessing the performances of various architectures and algorithms with imbalanced data have been carried out~\citep{japkowicz2002systematic,japkowicz2004svm,lemnaru2012systematic,buda2018cnn}, with a renewed interest in the era of big data and deep learning~\citep{ghosh2022deep,johnson2019deep}. Various strategies~\citep{anand1993improved,laurikkala2001improving,chawla2002smote,batista2004methods,he2008adasyn,alshammari2022tail}  based on restoring balance in the training set can be found in textbooks~\citep{he2013book,fernandez2018book}.
Despite this considerable body of work,
a theoretical understanding of learning with class imbalance remains elusive, which possibly impedes the design of optimal, task-adapted strategies.

In this work we propose a probabilistic setting that faithfully takes into account the imbalance ratio and essential features of the data, such as their first and second moments, in which the performances of linear classifiers can be analytically derived in the high-dimensional limit using the framework of the statistical mechanics of learning~\citep{engel2001book}. Based on this theory we can identify strategies to under or oversample the data in, respectively, the majority and minority classes offering optimal performances for the metrics of interest. 

\paragraph{Summary of the main results.} We report here the main results of our work.

In Sec.~\ref{sec:theo1}, we analytically characterize the asymptotic performances of linear classifiers trained on imbalanced datasets, when data points are discrete multi-state variables. These asymptotics are based on the so-called {\it replica method} from statistical physics. 

We provide {\it quantitative} behaviors of various metrics within this setting -- such as confusion matrix, accuracy (ACC), balanced accuracy (BA) and area under the ROC curve (AUC). This allows us to show, for example, that AUC is rather insensitive to imbalance, and is therefore not a reliable predictor of some aspects of the performances captured by more sensitive metrics, e.g. BA.

\begin{figure}[ht]
    \centering
    \includegraphics[width=0.48\textwidth]{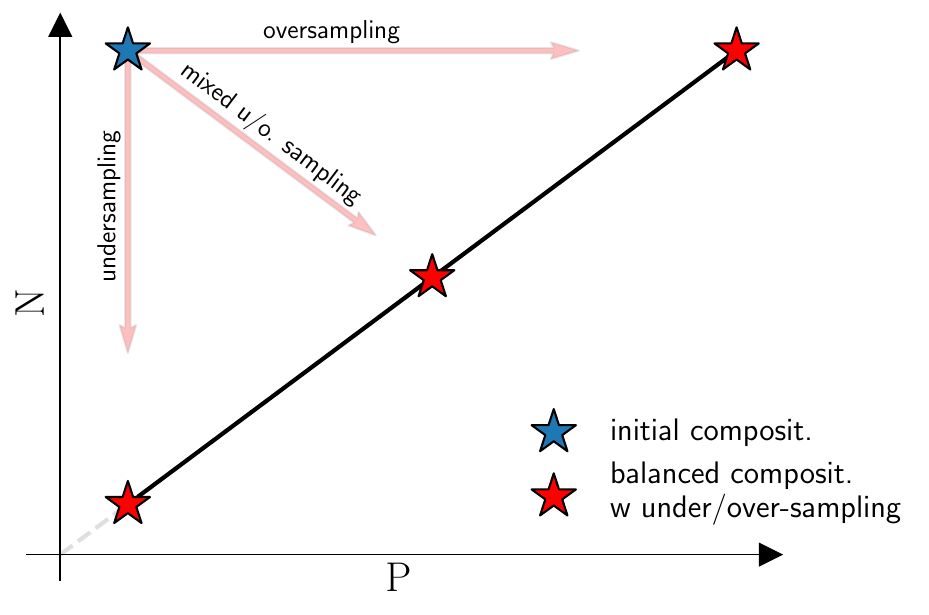}
    \vskip -.3cm
    \caption{Illustration of our restoring balance procedure. Classifiers trained on datasets with severe imbalance (blue star) generally show poor generalization performances. Restoring balance by mixing under and oversampling improves classification performances (red stars across the line $P=N$). Here $P,N$ indicate the sizes of the positive and negative classes, initially with $P\ll N$.}
    \label{fig:graph_abstract}
\end{figure}

In Sec.~\ref{sec:results}, we provide exact asymptotic predictions to assess the effect of simple under and/or oversampling strategies to mitigate the imbalance problem. We show that best performances are reached with mixed strategies restoring data balance (Fig.~\ref{fig:graph_abstract}).

In Sec.~\ref{sec:beyond}, we provide empirical support that our theoretical findings qualitatively hold with deeper classifiers and more sophisticated under/oversampling methods. In particular, we propose and test an unsupervised probabilistic model, learned on the minority class, to generate new minority-class instances or to filter out majority-class data (Fig.~\ref{fig:graph_abstract}).

\paragraph{Related works.} Learning imbalanced mixture models of data with linear classifiers has long been studied in the statistical physics literature \citep{delgiudice1989}. Recently, 
\citet{loureiro2021gmm,mignacco2020gmm,pesce2023} evaluated rigorously the generalization error of a linear classifier on data from Gaussian mixture models, on test sets distributed as the training set; in the present work, we consider instead data in the form of multi-state discrete variables not necessarily normally distributed, we compare different performance metrics and study the impact of mitigation strategies on the learning protocol. For rigorous results that go beyond the hypothesis of Gaussian data from this approach, see the recent~\citet{dandi2023universality}.
\citet{mannelli2023unfair} revisited this problem in the context of the fairness of AI methods; at variance with them, we consider a classification task with labels coinciding with class membership, and we focus on dataset-preprocessing mitigation strategies more than on algorithmic-based methods (loss-reweighting, coupled neural networks). The dynamics of learning with gradient-based methods in presence of class imbalance, which we do not address here, was recently studied by \citet{francazi2023theoretical}.
Other theoretical assessments from complementary point of views were formulated in the recent past to obtain optimal oversampling ratios for imbalanced datasets~\citep{shang2023yahoo,chaudhuri2023}.
An interesting comparison with our work is given by~\citet{menon2013statistical}, which proves that classifiers trained with empirically-balanced losses attain optimal performances in the limit of infinite data: as discussed in \cref{sec:sampling_strategies}, we can account for this setting within our formalism, but our work deals with the proportional asymptotics where both the input dimension and the training set size are large, for which we show that simple loss-reweighting methods are sub-optimal with respect to data augmentation techniques more sophisticated than random oversampling.

In addition, a plethora of methods have been proposed to effectively restore balance in datasets. SMOTE~\citep{chawla2002smote}, whose strategy is to oversample the minority class by linearly interpolating existing data points, is one of the most frequently applied algorithms and exists in the literature under more than 100 variations~\citep{kovacs2019emp}. The use of unsupervised generative model to first learn and then augment the minority class has been actively explored, \eg{} with generative adversarial networks~\citep{douzas2018gan}, autoencoders~\citep{mondal2023ae}, variational autoencoders~\citep{wan2017vae,ai2023vae}, etc. Our proposed approch is closely related to RBM-SMOTE, introduced by~\citet{zieba2015rbm}, where a Restricted Boltzmann Machine (RBM) is trained on the minority class and then used to generate new samples starting from intermediate points obtained by SMOTE. Similar restoring balance procedures have been showed in  \citet{mirza2021deep} to have a positive impact on different performances metrics.

\begin{table}[h!]
\caption{\label{tab:notations} Notations and conventions used in this work.}

\vskip 0.15in
\begin{center}
\begin{small}
    \begin{tabular}{lr}
    \toprule
        $L,Q$ & dimension of the data, size of the alphabet \\
        $P, N$ & positive and negative class sizes of training data \\
        $\alpha^{+}, \alpha^{-}$  & classes sizes scaled by data dimension \\
        $\rho^{+}, \rho^{-}$  & fractions of the two classes in the training dataset \\
        $\varphi^{+}, \varphi^{-}$  & fractions of the two classes in the test dataset \\
        $\langle \cdot \rangle_{\pm} $ & average over positive and negative data \\
        $\vb{M}, \bm{\delta}$ & midpoint and shift between the classes' centers\\
        $C$ & covariance matrix of the data \\
        $ \vb{J},b $ & weights and bias of the linear SVM \\
        $\kappa$  & margin of the linear SVM \\
        $\Delta^+$, $\Delta^-$ & pre-activations of the output neuron\\
        $\ell(y,\Delta)$ & training loss function\\
        $g(y,\Delta)$ & test loss function \\
        \bottomrule
    \end{tabular}
\end{small}
\end{center}
\vskip -0.1in
\end{table}

\section{Theoretical setting}
\label{sec:theo1}

We consider pairs $(\vb{v}^{\nu}, y^{\nu}) \in \mathbb{R}^{L \times Q} \times \{ -1,+1 \}$ with $\nu = 1, \dots, P+N $ of training data points independently sampled, where $L$ sets the dimension of data points and each symbol $v_i$ can take $Q$ values (Table~\ref{tab:notations}). For example, in image recognition tasks, $L$ is the number of pixels and $Q$ the number of possible colors of each pixel; in molecular biology, $L$ is the length of the biological sequence and $Q$ the size of the alphabet, {\em e.g.} 20 for amino acids.
In statistical physics, this kind of data are usually called Potts configurations; in computer science, they are routinely obtained via tokenization, for example in language models \citep{jurafsky2009speech} or image recognition~\citep{wu2020visual}. A convenient representation of these categorical variables is  one-hot encoding:  $s_{i}(t) = 1$ if $v_i=t$, 0 otherwise.

In particular, we focus on a binary classification task where the classes are labeled by $y = \pm 1$ and the training set is made of $P$ positive and $N$ negative examples, $\mathcal{T} = \{ (\vb{s}^\mu_+, y^\mu = +1)\}_{\mu=1}^P \cup \{ (\vb{s}^\nu_-, y^\nu = -1)\}_{\nu=1}^N$. The classes are \emph{imbalanced}, \ie{} in general $P \neq N$. We denote with $\alpha^{\pm}$ the classes' sizes scaled by the dimension of data, \eg{} $\alpha^+ = P/L$, and with $\rho^{\pm}$ the fractions of positive and negative examples in the training set, \ie{} $\rho^{\pm} = \alpha^{\pm}/(\alpha^{+} + \alpha^{-})$. 
We are interested in studying the statistical properties of the linear classifier 
\begin{equation}
    \hat{y} = f(\vb{s}) =  \sgn\Biggl[\sum_{i,t} \frac{J_i(t) s_i(t)}{\sqrt{L}}  - b \Biggr]\,,
    \label{eq:model}
\end{equation}
where the weights and bias $(\vb{J},b) \in \mathbb{R}^{L \times Q} \times \mathbb{R}$ define the model and are learnt over the training data. The scaling of the inner product with $\sqrt{L}$ is chosen to obtain components $J_{i}(t)\sim O(1)$ after regularization and training. Due to the property $\sum_t s_i(t)=1$ of one-hot encoding, a linear classifier described by the set of parameters $(J_i(t),b)$ is identical to the one defined by $(J_i(t)+u_i,b+\sum_i u_i/\sqrt{L})$ for any arbitrary real numbers $u_i$.
To avoid this overparametrization, we impose the zero-sum conditions $\sum_t J_i(t) = 0$ for all $i$.

Given the training set, the parameters of the model are learnt through the following Empirical Risk Minimization (ERM):
\begin{align}
    (\vb{J}^\star,b^\star) ={}& \underset{\vb J \in \mathcal{S},b \in \mathbb{R}}{\arg \min} \,\mathcal{L}(\vb J,b)\,,     \label{eq:minimization-Jb}\\
    \mathcal{L}(\vb J,b) ={}&  \sum_{\mu = 1}^P \ell \Bigl[ y^\mu=+1,\Delta^+_\mu (\vb J, b)\Bigr] \notag\\
    +{}& \sum_{\nu = 1}^N \ell \Bigl[ y^\nu=-1,\Delta^-_\nu (\vb J, b)\Bigr]
    ,     
    \label{eq:loss}\\
    \Delta^\pm_\mu (\vb J, b) ={}& \sum_{i,t} \frac{J_i(t) s_{\pm,i}^\mu(t)}{\sqrt{L}}  - b,   \label{eq:Deltas}
\end{align}
where $\ell: \{ -1,+1 \} \times \mathbb{R} \to \mathbb{R}^+$ is a loss function and
the set $\mathcal{S}$, over which the optimization problem~\eqref{eq:minimization-Jb} is defined, is
the surface of the $(L\times Q)$-dimensional sphere (spherical regularization) compatible with the zero-sum conditions. Notice that the variables $\Delta^\pm_\mu$ defined in Eq.~\eqref{eq:Deltas} are simply the pre-activations of the output neuron evaluated on the input points $\vb{s}^\mu_\pm$. A possible choice of the loss function, which we will use in the following as a case of study, is the hinge loss
\begin{equation}
    \ell ( y, \Delta) = \max (0, \kappa  - y \Delta )\,,
        \label{eq:hinge}
\end{equation}
for some positive parameter $\kappa$ called \emph{margin}. High values of $\kappa$ make the classifier less affected by noise in the training data, so that the learning protocol is more \emph{stable}.
The choice of the set $\mathcal{S}$ and of the hinge loss implies that the linear model~\eqref{eq:model} is a spherical perceptron with hinge loss~\citep{franz2019perceptron}, which is equivalent to a soft-margin support vector machine with L2 regularization, as detailed in the Appendix~\ref{app:SVM}. 

In this work we assume that the input data points belonging to the two classes are sampled independently from a distribution having the following first and second moments: 
\begin{equation}
    \begin{aligned}
        &\mean{s_i(t)}_\pm  = M_i(t) \pm \dfrac{\delta_i (t)}{2 \sqrt{L}}\,,\\
        &\mean{s_i(t)s_j(u)}_\pm - \mean{s_i(t)}_\pm \mean{s_j(u)}_\pm = C_{ij}(t,u)\,,
    \end{aligned}
    \label{eq:Data}
\end{equation}
where the angle brackets stand for the expectations over the positive and negative classes. The vector $\vb{M}$ represents the global center-of-mass of the positive and negative distributions, $\bm{\delta}$ is the (rescaled) shift between their centers. The scaling by the factor $1/\sqrt{L}$ with $\delta_i(t) \sim O(1)$ ensures that the classification task is non-trivial for a linear classifier in the high-dimensional regime $L\gg 1$. Notice that, to simplify the analysis, we are assuming the two classes to have the same covariance, a condition often referred to as {\it homoscedasticity}. 
Higher-order statistics of the data  is irrelevant for the asymptotic properties reported below  under mild  conditions on the cumulants \cite{monasson1992}.

\subsection{Statistical mechanics of the learning problem\label{sec:theorySM}}

The ERM problem introduced in the previous section can be rephrased in a statistical mechanics framework. We consider $(\vb{J},b)$ as the configuration of a `physical' system with energy $\mathcal{L}$.  The partition function of this system reads
\begin{equation}
    \Omega = \int \diff b \diff \mu(\vb J) \exp [ - \beta \mathcal{L}(\vb{J},b) ],
    \label{eq:Omega}
\end{equation}
where $\diff \mu(\vb J)$ is an opportune measure over the weights taking into account regularization and zero-sum conditions, as defined in Appendix~\ref{app:replicas}, Eq.~\eqref{eq:theoryJmeasure}. As the `inverse temperature' $\beta\to\infty$, the integral in Eq.~\eqref{eq:Omega} is dominated by the solution of the optimization problem~\eqref{eq:minimization-Jb}. Within this framework, we are able to derive the expected values of functions of $(\vb{J}^\star,b^\star)$ over the data. In particular, we can characterize the asymptotic performance of the linear classifer. 

\begin{proposition}
\label{propERM}
Consider the ERM problem in Eq.~\eqref{eq:minimization-Jb} under the assumption of training data distribution \eqref{eq:Data}. Let the fractions $\varphi^{\pm}$ be the composition of the test set and 
\begin{equation}
    \mathcal{M}_{\mathrm{gen.}}(\varphi^+, \varphi^-) = \sum_{y \in \{\pm \} } \varphi^y \braket{  g ( \Delta^y,  y ) }_{\Delta^y},
    \label{eq:generalization_metrics}
\end{equation}
a performance metric of choice, with $\Delta^{\pm}$ defined as in Eq.~\eqref{eq:Deltas} over test data points and $g$ a generic test loss function (see~\cref{tab:metrics} for examples). In the asymptotic regime 
$ L, P, N \to \infty\ \ \text{at fixed ratios}\ \  
\alpha^+ = P/L$ and $ \alpha^- = N/L $, the expected value in Eq.~\eqref{eq:generalization_metrics} can be taken over the normal distribution 
\begin{equation}
    p(\Delta^{\pm})  = \mathcal{N}\left( \pm \dfrac{r}{2} - b, q \right),
    \label{eq:DeltaDistr}
\end{equation}
whose parameters $q,r,b$ are the solution of the saddle-point equations stemming from the extremization of the following quantity
\begin{equation}
    \mathbb{E}_\mathrm{data} \left[ \log \Omega \right] \sim \frac{\beta L}{2} \extr_{\substack{b,r,q,x\\\hat{k},\hat{r},\hat{q},\hat{x}}} ( G_S+\alpha^+ G_+ + \alpha^- G_-).
    \label{eq:logOmega}
\end{equation}
\end{proposition}

In the above proposition -- fully derived in Appendices~\ref{app:replicas}.1-4, \ref{app:metrics} -- the functions $G_\pm$ and $G_S$ in Eq~\eqref{eq:logOmega} are defined as
\begin{multline}
    G_\pm 
    = -\sqrt{\frac{q}{2\pi }} \left( \frac{x - \mathcal{K}_\pm }{x} e^{-\frac{(x - \mathcal{K}_\pm)^2}{2 q}}+ \frac{\mathcal{K}_\pm}{x}  e^{-\frac{\mathcal{K}_\pm^2}{2 q}}
      \right) \\   
    +\frac{(x - \mathcal{K}_\pm)^2+q}{x} H\!\left(\frac{x-\mathcal{K}_\pm}{\sqrt{q}}\right)
    -\frac{\mathcal{K}_\pm^2+q}{x} H\!\left(\frac{-\mathcal{K}_\pm}{\sqrt{q}}\right)\!,\!\!\!\! \label{eq:GpmUNSAT}
\end{multline}
and
\begin{multline}
    G_S = Q \hat{k}  +\hat{x} q   +\hat{q} x + 2\hat{r} r  +  \frac{\hat{r}^2}{L} \bm{\delta}^\top A^{-1} \bm{\delta}\\
    - \frac{\hat{q}}{L} \tr(A^{-1} C) ,
 \label{eq:GSUNSAT}
\end{multline}
where $\mathcal{K}_\pm = \kappa - r/2 \pm b$, $ H(x) = \erfc(x/\sqrt{2})/2$ ($\erfc$ is the complementary error function), $A = \hat{k} \mathbb{I}_Q\otimes \mathbb{I}_L + \hat{x} C$ ($\otimes$ is the Kroenecker product). Here, algebraic operations as traces and vector/matrix multiplications are performed in the $(L\times Q)$-dimensional linear space spanned by the Potts and position indices. 

We stress here that Proposition~\ref{propERM} holds under the hypotheses that higher cumulants in the distribution of the data can be neglected in Eq.~\eqref{eq:logOmega} for large $L$ with respect to the first and second cumulants reported in Eq.~\eqref{eq:Data}, and that the optimization problem~\eqref{eq:minimization-Jb} is strictly convex; the hypothesis of strict convexity is inessential and can be relaxed to convexity, in which case Eqs.~\eqref{eq:GpmUNSAT} and~\eqref{eq:GSUNSAT} need to be substituted with~\eqref{eq:appGpmSAT} and~\eqref{eq:appGS}, as discussed in Appendix~\ref{app:replicas}.
The extremization in Eq.~\eqref{eq:logOmega} is performed numerically with respect to the bias $b$, the \emph{order parameters} $x$, $q$, $r$, and the conjugate variables $\hat{x}$, $\hat{q}$, $\hat{r}$, $\hat{k}$. This procedure provides our analytical predictions for training and generalization metrics as a function of $\alpha^+,\alpha^-$ and the data statistics, $\vb{M}$, $\bm{\delta}$ and $C$. The order parameters, whose precise definition in terms of replica-symmetric overlap matrices is detailed in the Appendix, have a simple physical interpretation in terms of the low-order statistics of the variables $\Delta^{\pm}_{\text{test}}(\mathbf{J}^\star,b^\star)$, defined over \emph{test} data points. 
The order parameters can also be interpreted in geometric terms. In particular, $r$ measures how well $\vb{J}^\star$ is aligned along the vector $\bm{\delta}$, $q$ measures the inner product $(\mathbf{J}^\star)^\top C \mathbf{J}^\star /L$, and $b$ is the optimal value of the bias in Eq.~\eqref{eq:model}. 

Detailed calculations and more general results, such as the expression for the functions $G_S$, $G_\pm$ in the linearly separable phase of the classifier, are reported in Appendix~\ref{app:replicas}.2-3.

\subsection{Performance metrics\label{sec:metrics}}

\begin{table}[t]
\caption{Analytical predictions for the elements of the confusion matrix and common performance metrics derived from them. Brackets indicate averages w.r.t. the distributions~\eqref{eq:DeltaDistr}, the threshold $\gamma$ is introduced to define the ROC and the PR curves, $\varphi^\pm$ indicates the composition of the test set and $\theta$ the Heaviside step function. Other metrics are reported in Appendix.}
\label{tab:metrics}
\vskip 0.15in
\begin{center}
\begin{small}
    \begin{tabular}{lcr}
    \toprule
        $\text{TPR}(\gamma)$  & $ \braket{\theta(\Delta^+ - \gamma)}_{\Delta^+} $ & True positive rate\\ \addlinespace
        $\text{FPR}(\gamma)$  & $ \braket{\theta(\Delta^- - \gamma)}_{\Delta^-} $ & False positive rate\\ \addlinespace
        $\text{TNR}(\gamma)$  & $ 1 - \text{FPR}(\gamma) $ & True negative rate\\ \addlinespace
        $\text{FNR}(\gamma)$  & $ 1 - \text{TPR}(\gamma) $ & False negative rate\\ \addlinespace
        $\text{PPV}(\gamma)$  & $ \dfrac{\varphi^+ \text{TPR}(\gamma)}{\varphi^+ \text{TPR}(\gamma) + \varphi^- \text{FPR}(\gamma)}$ & Precision\\ \addlinespace
        $\text{ACC}$  & $ \varphi^+ \text{TPR}(0) + \varphi^- \text{TNR}(0) $ & Accuracy\\ \addlinespace
        $\text{BA}$  & $ [\text{TPR}(0) + \text{TNR}(0)]/2 $ & Balanced accuracy\\ \addlinespace
        $\text{AUC}$  & $ \int_{-\infty}^{\infty}\! \diff \gamma | \text{FPR}'(\gamma)| \text{TPR}(\gamma)$ & Area under ROC\\ \addlinespace
        $\text{AUPRC}$  & $ \int_{-\infty}^{\infty}\! \diff \gamma | \text{TPR}'(\gamma)| \text{PPV}(\gamma)$ & Area under PRC\\
        \bottomrule
    \end{tabular}
\end{small}
\end{center}
\vskip -0.1in
\end{table}

The approach above, and in particular Eq.~\eqref{eq:DeltaDistr} evaluated on the values of the order parameters obtained from the extremization~\eqref{eq:logOmega}, provides analytical expressions for the generalization performance, assessed through the elements of the confusion matrix from Table~\ref{tab:metrics}.  For example, the accuracy $\text{ACC}(\varphi^+, \varphi^-)$ (or 0/1 accuracy) referred to as ACC in Table~\ref{tab:metrics} is obtained from Eq.~\eqref{eq:generalization_metrics} with $g(y,\Delta^y)=\theta(y\, \Delta^y)$, where $\theta$ is the Heaviside step function.
The choice $\varphi^\pm = 0.5$ corresponds to the \emph{balanced accuracy} (BA), a popular measure of performance with imbalanced datasets, while $\varphi^+ = \rho^+$ stands for test as imbalanced as training.

The expression of the balanced accuracy BA is generally involved, but simplifies for large margin $\kappa \gg 1$. In this regime, we obtain explicit rates of convergence in terms of the training set size $(\alpha^+, \alpha^-)$. In the case of $Q=2$ states ($t=\pm$), position-independent $M_i(t=\pm) = (1\pm m)/2$ and diagonal covariance matrix $C_{ij}(t=+,u=+) = \delta_{ij} (1-m^2)/4$, we obtain the peak value of the BA at $\alpha^+ = \alpha^-$ as 
\begin{equation}
    \text{BA}(\alpha^+ = \alpha^-) \sim 1 - \dfrac{1}{2}  \erfc\! \left( \dfrac{\lVert\bm{\delta} \rVert_2}{2\sqrt{2L(1-m^2)}}\right) + \dfrac{\mathcal{C}}{2\alpha^+},
    \label{eq:misclassification_error_largeK}
\end{equation}
where $\mathcal{C}$ is a constant term, while $\text{BA}(\alpha^+ \neq \alpha^-) = 0$.

\section{Theoretical results on imbalanced learning\label{sec:results}}   

\subsection{Behaviour of common performance metrics\label{sec:res_metrics}}

In this section we analyze the metrics defined in Table~\ref{tab:metrics} based on the theory explained above. We first notice, from their definition, that these metrics are impacted by imbalance both \emph{explicitly}, due to the dependence on the ratios $\varphi^\pm$ expressing the composition of the test set, and \emph{implicitly}, due to the fact that they are evaluated on a model trained on an imbalanced dataset. Common choices for the test set are $\varphi^+ = \rho^+$ (test set distributed as the training set) or $\varphi^+ = 0.5$ (balanced test set).

We report in Fig.~\ref{fig:fig_theory}a the analytical curves as a function of the training imbalance ratio $\rho^+$, obtained in the theoretical setting explained above, for $\varphi^+ = \rho^+$. We conclude that:
\begin{itemize}
    \item ACC, even if exhibiting a peak around the balanced point $\rho^+ = 0.5$, predicts best performances when the dataset is heavily imbalanced. This is expected: due to the explicit dependence on $\varphi^\pm$ and the fact that $\varphi^+ = \rho^+$,  ACC only evaluates how well the majority class is predicted, even though the model behaves as a random classifier on the minority class.
    \item BA, which has no explicit dependence on the composition of the test set (it weights in the same way the probabilities of true label prediction on the majority and minority classes), assign best performances to the model trained on a balanced dataset.
    \item AUC predicts best performances away from the balanced point.  AUC is defined as the area under the receiver operating characteristic (ROC) curve, that is the parametric curve $(\text{FPR}(\gamma),\text{TPR}(\gamma))$ as a function of the threshold on the predictor $\gamma$. 
    As $\gamma\in(-\infty,\infty)$, AUC is independent from the bias $b$ in Eq.~\eqref{eq:DeltaDistr}, which is the observable most impacted by imbalance (see its trend obtained from our theory in Appendix~\ref{app:replicas}, Fig.~\ref{fig:app_metrics_behavior}, and many observations in literature, \eg~\citet{chaudhuri2023}).
    \item AUPRC, \ie{} the area under the Precision-Recall curve (PRC -- the parametric curve $(\text{TPR}(\gamma),\text{PPV}(\gamma))$ as a function of $\gamma$), is often proposed as a better alternative to AUC in presence of imbalance. As AUC, it does not depend on the bias $b$, and thus does not  directly measure the parameter most sensitive to imbalance. However, due to the explicit dependence on the test set (notice that $\text{PPV}(\infty) = 1$, $\text{PPV}(-\infty) = \varphi^+$), AUPRC is always bounded from below by $\varphi^+ = \rho^+$, increasing ``by definition'' as the positive class ratio in the dataset.
\end{itemize}
For more of these metrics, we address the reader to Appendix~\ref{app:metrics} and Table~\ref{tab:app_metrics}. We conclude from this analysis that, for different reasons we can explain within our theoretical setting, standard generalization metrics can be misleading about the performance of a model trained under imbalance. For the rest of this paper we will mostly make use of BA, which is one of the most robust in predicting best performances around the balanced point.

\subsection{Agreement with results on real datasets.}
We show in \cref{fig:fig_theory}b that the theory devised in this work is able to \emph{quantitatively} predict the BA curves (as well as other generalization curves) of linear SVMs trained on standard benchmark datasets. In particular, we validated our predictions on (i) the Parity MNIST (pMNIST) dataset, having odd and even digits classes; (ii) Fashion MNIST (FMNIST) with classes containing "Pullover" and "Shirt" images; and (iii) CelebA with classes containing faces with "Straight hair" and "Wavy hair". Note that our theory assumes the two classes in the training dataset to have the same covariance matrices: while this is not generally true on all datasets, we observe agreement between numerical experiments and asymptotic predictions, by feeding the mean empirical covariance of positive and negative data, $C = (C^+_{\text{emp}} + C^-_{\text{emp}})/2$, in our theory. Such agreement remains quantitatively valid provided that the covariance matrices $C^{+}_{\text{emp}}$, $C^{-}_{\text{emp}}$ are similar, and is only qualitative for more diverse covariances (\eg{} 0/1 MNIST). Moreover, we report in Appendix \ref{sec:app_LP} a benchmark dataset with Lattice Proteins where diagonal covariances are sufficient to predict quantitatively the numerical results.

\subsection{Balance-to-performance trade-off\label{sec:res_tradeoff}}

\begin{figure*}[ht]
    \centering    \includegraphics[width=\textwidth]{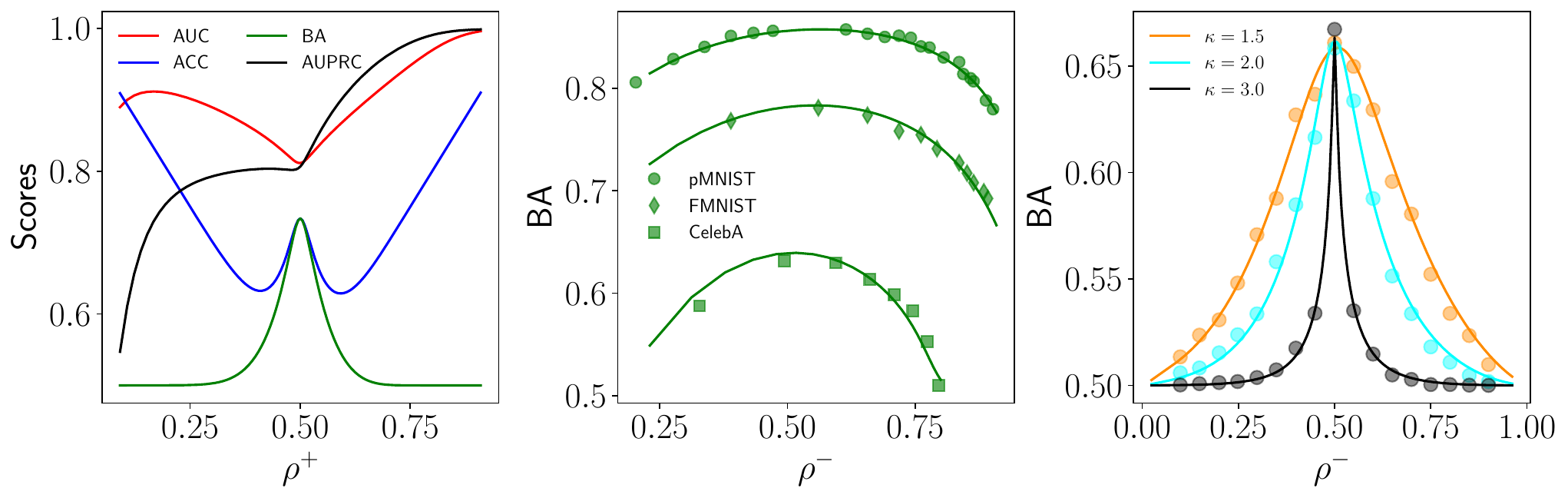}
    \put(-485,160){{\bf a}}
    \put(-325,160){{\bf b}}
    \put(-165,160){{\bf c}}
    \vskip -.3cm
    \caption{\label{fig:fig_theory}
 Analytical results derived within our framework.  {\bf a)} Different metrics on synthetic data evaluated on a test set having same train set composition. Here $L=100$, $\kappa =2$, $\alpha^{-}=5$ with $C, \delta$ sampled randomly. {\bf b)} Analytical predictions for benchmark datasets (MNIST, FashionMNIST and CelebA). Dots show numerical simulations, averaged over $10$ trials for $\kappa =0.5$. {\bf c)} Balanced accuracy curves as a function of $\rho^-$. The margin $\kappa$ of the algorithm controls the balance-to-performance trade-off. Dots correspond to numerical simulations with $\alpha^+ = 2$, $Q=2$, $L=100$, position-independent $\vb{M}$, normally-distributed $\bm{\delta}$, and diagonal covariance $C$ averaged over $50$ trials. }
\end{figure*}

Using analytical results from Sec.~\ref{sec:theo1}, we can assess to what extent the imbalance of the dataset and the stability of the classifier impact the performance. We evaluate results in terms of the BA for different compositions of the training set. 
The generalization curves in \cref{fig:fig_theory}c exhibit the presence of a balance-to-performance trade-off: for high values of $\kappa$, the accuracy is extremely sensitive to the composition of the training set, and the outputs of the classifier are essentially random unless $\rho^- = \rho^+ \sim 0.5$. Increasing the margin $\kappa$ results in an absolute gain of accuracy as a function of $\rho^-$, but narrows down the window outside of which the classifier acts as a random one.

We can get intuition on the existence of such trade-off by looking at the high-$\kappa$ and $\rho^+ = 0.5$ case. In this regime, the bias of the algorithm $b$ is then factoring out the center-of-mass $\vb{M}$ of the data, while the weight vector $\vb{J}$ becomes fully aligned with the displacement $\bm{\delta}$ between the two classes, yielding the optimal solution for the classification task. However, as soon as one moves away from the balanced training set, the extremely steep behaviour of the parameters $\vb{J}$ and $b$ around $\rho^+ = \rho^- = 0.5$ -- due to the high margin -- makes the performance drop down (see \cref{fig:app_metrics_behavior}a  in the Appendix \ref{app:replicas}). For milder values of $\kappa$, the effect is less pronounced, even though $\vb{J}$ and $b$ are not optimal.

For real data, the value of the margin $\kappa$ of the linear classifier has to be  compared to the data statistics to determine how tight the balance-to-performance trade-off is. Theory suggests that what matters is an effective margin depending on the ratio $\eta (C) /{\lVert \bm{\delta} \rVert}_2^2$,
where $\eta(C)$ denotes the largest eigenvalue of the covariance matrix.
We check this on pMNIST and CelebA datasets, where the ratio $\eta(C)/\lVert \bm{\delta} \rVert_2^2$ is larger for CelebA data and thus results in a tighter good performance window (see \cref{fig:fig_theory}b).

Hence, from the interplay between $\kappa$ and data statistics and composition arises the balance-to-performance trade-off.

\subsection{Insights on standard over vs. undersampling}
\label{sec:sampling_strategies}
The theoretical framework of Sec.~\ref{sec:theo1} allows us to gain some understanding on  {\it whether it is better to restore balance by oversampling the minority class or undersampling the majority class}. With no loss of generality we assume the minority class to be the positive one; hence, the negative class size is $\alpha^- >\alpha^+$.
\begin{figure}[ht]
    \centering
    \includegraphics[width=0.49\textwidth]{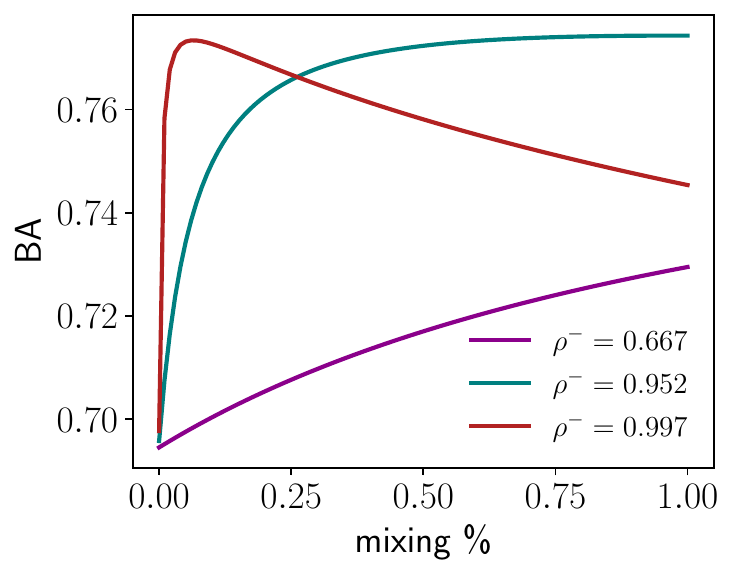}
    \vskip -.4cm\caption{\label{fig:fig_mixing} Optimal mixing strategy. We report theoretical predictions for the BA metric as a function of the mixing under/oversampling percentage in the training set. Depending on the initial training set composition $(\rho^+, \rho^-)$, one can select the optimal strategy to restore balance. Here $L=100$, $\kappa =0.5$, with $C$ and $\bm{\delta}$ sampled randomly.}
\end{figure}

The full undersampling strategy consists in randomly removing negative data points down to $\alpha^- = \alpha^+$, a strategy called Random Undersampling (RUS). To  oversample positive data, we modify Eq.~\eqref{eq:loss} by introducing a factor $c_{\mu}\ge  1$ in front of the loss function for positive samples, accounting for their multiplicity due to duplication. For the sake of simplicity, we assume $\left< c_{\mu} \right> = c \in \mathbb{R}$, \ie{} we uniformly sample each positive data $c$ times, with $ c\in \left[1, \alpha^- /\alpha^+ \right]$. 
The limit case $c=1$ corresponds to leaving the dataset as it is (RUS), while when $c = {\alpha^-}/{\alpha^+}$ the minority class is duplicated up to the majority class size, a strategy known as Random Oversampling (ROS). In this respect, ROS is theoretically equivalent to a simple loss-reweighting strategy that scales the contribution of each class to the total loss by the inverse of their size. Intermediate strategies mixing under and oversampling are associated to intermediate values of $c$. Based on our observation from the previous section, we study only the case of balanced training set looking at BA -- even though our theory allows to make prediction for any training set composition and metrics. Thus we define the mixing percentage as 
\begin{equation}
    \text{mixing \%} = \dfrac{\alpha^{+}}{\alpha^{-} - \alpha^{+}} (c-1),
\end{equation}
and we ask what is the optimal mixing percentage in terms of BA, \ie{} what is the best $c$ value?

In \cref{fig:fig_mixing} we observe that for all tested cases, full undersampling is a sub-optimal technique among restoring balance protocols, while mixed under/oversampling and oversampling should be preferred. In particular, for severe imbalance a mixed strategy gives highest BA compared to full oversampling, which should be preferred when the imbalance is milder: this is also numerically observed on real data in \cref{fig:fig_numexps}b, dashed lines. However, we expect that in general the optimal strategy strongly depends not only on the imbalance ratio between classes, but on absolute dataset size, on classes' similarity in feature space and on the margin $\kappa$ used in the linear classifier.

In this respect, our theoretical framework can be useful as it offers a quantitative way to evaluate {\it a priori} the performances based on the first- and second-order statistic of the data, thus providing a guideline in practical applications.

\section{Numerical experiments\label{sec:beyond}}
In this section we investigate with numerical experiments if the phenomenological findings at the theoretical level hold when we go beyond the limits of our theory. Specifically, we developed a framework for liner SVMs and for ROS/RUS techniques. We now ask what happens with deeper neural networks and more involved under/oversampling techniques. We address these two questions separately, to factor out any other element. 

\subsection{Improved sampling strategies}
First, we examine the effect of sampling strategies more involved than random sampling, resorting to Restricted Boltzmann Machines (RBMs) that allow to under and oversample at the same time with a protocol we call Likelihood-Informed Sampling (LIS). We train the model on the positive (minor) class solely and use it to generate new positive digits and to subsample negative digits based on their likelihood. A closely related approach to generate new examples with RBMs has been introduced in~\citet{zieba2015rbm}.

\textbf{Restricted Boltzmann machines.} RBMs are bipartite graphical models, and include $L$ visible units $\vb{v} = (v_1, \cdots , v_L )$ and $M$ latent (or hidden) units $\vb{z} = (z_1, \cdots, z_M)$. Only connections between visible and latent units are allowed through the interaction weights $w_{i \mu}$. RBMs define a joint probability distribution over $\vb{v}$ and $\vb{z}$ as the Gibbs distribution 
\begin{multline}
    p(\vb{v},\vb{z}) = \dfrac{1}{Z} \exp \left\{  \sum_{i=1}^L h_i(v_i) - \sum_{\mu=1}^M U_{\mu} (z_{\mu}) \right. \\ \left. + \sum_{\mu=1}^M \sum_{i=1}^L w_{i \mu} (v_i) z_{\mu}
    \right\}, 
\end{multline}
where the $h_i$'s and $U_{\mu}$'s act on, respectively, visible and latent units, and the last term couples the latent and visible layers. Learning of the model parameters is achieved by maximizing the marginal likelihood $p(\vb{v}) \equiv\int \diff{\vb{z}} \,p(\vb{v},\vb{z})$ over the training set. Details about the numerical implementation can be found in Appendix~\ref{app:num_exps}. After training and validation, the model can be used to score any new data $\vb{v}'$ through $\log p(\vb{v}')$.

\textbf{Likelihood-informed sampling. \label{sec:LIS} } We use RBM scores to form an appropriate balanced dataset by oversampling the minority class or undersampling the majority one, as follows.

\begin{enumerate}[(i)]
    \item for oversampling, we rely on the generative power of our unsupervised architecture. Starting from an initial configuration, we perform Gibbs sampling in data space based on the model scores. We retain only samples having scores similar to the ones in the minority data class, as low-score samples are considered not representative of the class. 
    \item for undersampling, we filter out majority class samples at the boundaries of the log-likelihood distribution, as they are potentially less informative to the classification task.
\end{enumerate}

\textbf{Results.} We perform numerical experiments on MNIST (classes with smaller and larger than digit 5) and with linear SVMs. Using RBMs for LIS we show that there is an improvement of performances as it was happening for ROS/RUS, even when starting from low positive class sizes; moreover, this sampling strategy yields better results compared to naive random under/oversamplings (see \cref{fig:fig_numexps}a). \\
The effect of restoring balance of data can be seen through a geometric visualization of the problem. We train separately two linear SVMs with imbalanced ERM and with ERM after restoring balance  between the two classes of the training set. We then project the test set data in the 2d space defined by the two optimal decision boundaries found by the classifier (see \cref{fig:fig_numexps}b). We observe that data balance between classes allows the model to place a more effective decision boundary (gold line), while imbalanced ERM (purple line) only learns the majority class.

\subsection{Deep classifiers}
\begin{figure*}[ht]
    \centering
        \includegraphics[width=\textwidth]{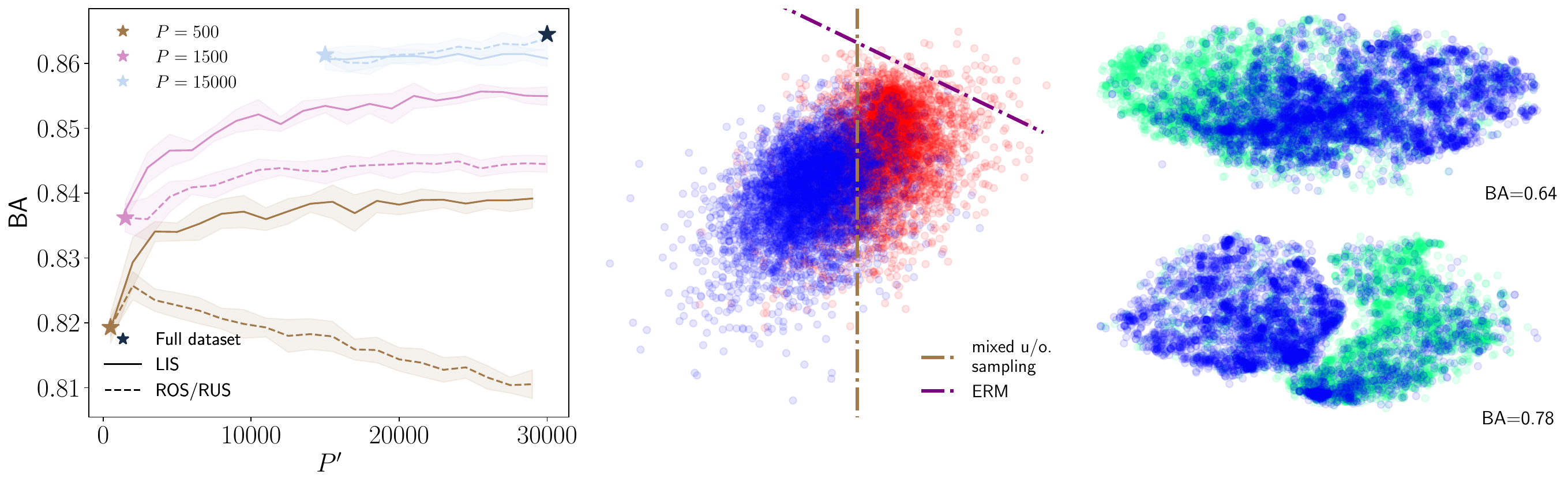}
    \put(-480,160){{\bf a}}
    \put(-285,160){{\bf b}}
    \put(-150,160){{\bf c}}
    \vskip -.5cm
    \caption{ \label{fig:fig_numexps} Numerical investigation on improved sampling techniques and deep classifier ResNet-50. {\bf a)} Mixed sampling strategies to obtain a balanced training set for classification with a linear SVM on binary MNIST, as a function of the new sample size $P^{\prime}$. As random sampling techniques, also higher level methods lead to an increase of performance. {\bf b)} Geometrical interpretation of the effect of restoring balance (gold line) on the decision boundary of a linear SVM compared to imbalanced ERM (purple line). Data points are the MNIST test set. {\bf c)} We visualize test data classification in the last feature layer of the network through tSNE, for imbalanced and balanced training set (top and bottom, respectively). The network trained on balanced data achieves improved performances, separating better the two classes.} 
\end{figure*}

We consider an ImageNet pretrained ResNet-50 and finetune it on a binarized version of Cifar10 (\ie{} images are split in two classes -- positive and negative -- depending if their label is smaller or larger than 5). To understand if the imbalance plays a relevant role similarly to the case with linear SVMs, we run two parallel experiments with imbalanced and balanced training set, $P=0.1N$ and $P=N$ respectively. All network parameters in the two experiments stay the same. We show that the balanced experiments achieves higher accuracy by looking at the last feature layer of the network (see \cref{fig:fig_numexps}c).

\section{Discussion and perspectives}

In this work we devise a theoretical framework to study the generalization performance of linear models for binary classification tasks under imbalanced training datasets. We show which metrics are more informative than others in imbalance learning and we give sharp estimations on the optimal mixing of under and oversampling strategy given the data statistics. We extend our study beyond the limit of the theoretical analysis, supporting our findings with numerical investigations. 

To the best of our knowledge, this is the first asymptotic characterization of linear classifiers trained over imbalanced datasets for any generalization metrics and categorical data. Looking forward, we believe that while restoring data balance techniques are unanimously empirically beneficial for ERM, their theoretical understanding falls behind, and our work can help closing this gap. 
In this regard, our study can be extended along several directions. First, our theoretical setting assumes that the two classes of data differ by their first-order statistics while share the same second-order one. We would like to pursue this direction and derive analytical predictions for {\it heteroscedastic} classes, motivated by the stream of results obtained for high-dimensional data with non-trivial structures, such as correlated patterns~\citep{monasson1992,lopez1995storage}, Gaussian covariate and mixture models~\citep{loureiro2021gmm}, random features models~\citep{goldt2020gep}, object manifolds~\citep{chung2018manifold}, simplex learning~\citep{pastore2020structure,rotondo2020beyond,pastore2021critical,baroffio2024resolution}. 

Moreover, it would be interesting to investigate, in light of our findings, the behavior of deep classifiers: how does the sensitivity to class imbalance scale with the expressive power of the architecture? For a recent review on the phenomenology, see~\citet{ghosh2022deep}. While deep learning models in general remain analytically intractable, theoretical understanding has been reached in certain asymptotic regimes (mainly, the infinite-width limit and the proportional scaling between width and size of the training set), both for fully connected~\citep{jacot2018NTK,lee2018gaussian,pacelli2022finitewidth,cui2023optimal,cui2024asymptotics} and convolutional architectures~\citep{naveh2021,aiudi2023local}. 

Analysis in a framework similar to the one we provide has been proven to work not only for linear models, but also for kernel machines~\citep{dietrich1999svm,gerace2020generalisation,bordelon2020spectrum,aguirrelopez2024random}. We expect our approach to be easily generalizable to these cases, by simply replacing the input space we consider here with the feature space of those models, and the input statistics (means and covariances) with the one of the features (means of the features, kernel in feature space). Deep neural networks have been proven equivalent to kernel machines in certain cases, namely the infinite-width limit mentioned above; in this sense, kernels provide lower bounds on the performance of deep neural networks that work in more generic regimes of feature learning. Therefore, our findings, which are sharp for linear models and could be extended to kernel machines, can provide insights on generalization also for deep neural networks.
Considering the large use of deep learning in practical applications, a theoretical assessment of its properties in the case of imbalanced data is very much needed.

Lastly, our theory could be extended beyond binary classification as in several applications one faces imbalance within the context of multiclass classification tasks.

As for numerical experiments, we introduced a strategy to restore balance based on unsupervised probabilistic models proving that it improves performance compared to naive imbalanced ERM.  At a speculative level, RBMs have the potential to capture interpretable features of the minority class, while saving computational cost compared to deeper architectures. This restoring balance approach could also benefit from recent progress on efficient out-of-equilibrium sampling with RBMs~\citep{agoritsas2023ebm,carbone2023fast}. We believe that similar increase in  performances would follow from restoring data balance with other methods~\citep{mirza2021deep}.
Geometrically, we showed that restoring balance allowed one to remove the bias of the decision boundary towards the majority class only, as also observed in~\citet{chaudhuri2023}.

\section*{Impact Statement}
This paper aims to advance the field of Machine Learning. Class imbalance in datasets represents a common issue in many Machine Learning applications, since it undermines performance and makes predictions hard to assess with adequate metrics. We consider crucial to have a theoretical control on how imbalance impacts learning, and how it can be alleviated.

\section*{Acknowledgements}
We acknowledge funding from the CNRS - University of Tokyo \textit{``80 Prime''} Joint Research Program and from the Agence Nationale de la Recherche (ANR-19 Decrypted CE30-0021-01 to S.C. and R.M.).
\bibliography{refs}

\begin{thebibliography}{81}
\providecommand{\natexlab}[1]{#1}
\providecommand{\url}[1]{\texttt{#1}}
\expandafter\ifx\csname urlstyle\endcsname\relax
  \providecommand{\doi}[1]{doi: #1}\else
  \providecommand{\doi}{doi: \begingroup \urlstyle{rm}\Url}\fi

\bibitem[Agoritsas et~al.(2018)Agoritsas, Biroli, Urbani, and
  Zamponi]{agoritsas2018cavity}
Agoritsas, E., Biroli, G., Urbani, P., and Zamponi, F.
\newblock Out-of-equilibrium dynamical mean-field equations for the perceptron
  model.
\newblock \emph{Journal of Physics A: Mathematical and Theoretical},
  51\penalty0 (8):\penalty0 085002, Jan 2018.
\newblock \doi{10.1088/1751-8121/aaa68d}.
\newblock URL \url{https://dx.doi.org/10.1088/1751-8121/aaa68d}.

\bibitem[Agoritsas et~al.(2023)Agoritsas, Catania, Decelle, and
  Seoane]{agoritsas2023ebm}
Agoritsas, E., Catania, G., Decelle, A., and Seoane, B.
\newblock Explaining the effects of non-convergent {MCMC} in the training of
  energy-based models.
\newblock In Krause, A., Brunskill, E., Cho, K., Engelhardt, B., Sabato, S.,
  and Scarlett, J. (eds.), \emph{Proceedings of the 40th International
  Conference on Machine Learning}, volume 202 of \emph{Proceedings of Machine
  Learning Research}, pp.\  322--336. PMLR, 23--29 Jul 2023.
\newblock URL \url{https://proceedings.mlr.press/v202/agoritsas23a.html}.

\bibitem[Aguirre-López et~al.(2025)Aguirre-López, Franz, and
  Pastore]{aguirrelopez2024random}
Aguirre-López, F., Franz, S., and Pastore, M.
\newblock {Random features and polynomial rules}.
\newblock \emph{SciPost Phys.}, 18:\penalty0 039, 2025.
\newblock \doi{10.21468/SciPostPhys.18.1.039}.
\newblock URL \url{https://scipost.org/10.21468/SciPostPhys.18.1.039}.

\bibitem[Ai et~al.(2023)Ai, Wang, He, Wen, Pan, and Xu]{ai2023vae}
Ai, Q., Wang, P., He, L., Wen, L., Pan, L., and Xu, Z.
\newblock Generative oversampling for imbalanced data via majority-guided vae.
\newblock In Ruiz, F., Dy, J., and van~de Meent, J.-W. (eds.),
  \emph{Proceedings of The 26th International Conference on Artificial
  Intelligence and Statistics}, volume 206 of \emph{Proceedings of Machine
  Learning Research}, pp.\  3315--3330. PMLR, 25--27 Apr 2023.
\newblock URL \url{https://proceedings.mlr.press/v206/ai23a.html}.

\bibitem[Aiudi et~al.(2025)Aiudi, Pacelli, Baglioni, Vezzani, Burioni, and
  Rotondo]{aiudi2023local}
Aiudi, R., Pacelli, R., Baglioni, P., Vezzani, A., Burioni, R., and Rotondo, P.
\newblock Local kernel renormalization as a mechanism for feature learning in
  overparametrized convolutional neural networks.
\newblock \emph{Nature Communications}, 16\penalty0 (1):\penalty0 568, Jan
  2025.
\newblock ISSN 2041-1723.
\newblock \doi{10.1038/s41467-024-55229-3}.
\newblock URL \url{https://doi.org/10.1038/s41467-024-55229-3}.

\bibitem[Akbani et~al.(2004)Akbani, Kwek, and Japkowicz]{japkowicz2004svm}
Akbani, R., Kwek, S., and Japkowicz, N.
\newblock Applying support vector machines to imbalanced datasets.
\newblock In Boulicaut, J.-F., Esposito, F., Giannotti, F., and Pedreschi, D.
  (eds.), \emph{Machine Learning: ECML 2004}, pp.\  39--50, Berlin, Heidelberg,
  2004. Springer Berlin Heidelberg.
\newblock ISBN 978-3-540-30115-8.
\newblock \doi{10.1007/978-3-540-30115-8_7}.
\newblock URL \url{https://doi.org/10.1007/978-3-540-30115-8_7}.

\bibitem[Alshammari et~al.(2022)Alshammari, Wang, Ramanan, and
  Kong]{alshammari2022tail}
Alshammari, S., Wang, Y.-X., Ramanan, D., and Kong, S.
\newblock Long-tailed recognition via weight balancing.
\newblock In \emph{2022 IEEE/CVF Conference on Computer Vision and Pattern
  Recognition (CVPR)}, pp.\  6887--6897, 2022.
\newblock \doi{10.1109/CVPR52688.2022.00677}.
\newblock URL \url{https://doi.org/10.1109/CVPR52688.2022.00677}.

\bibitem[Anand et~al.(1993)Anand, Mehrotra, Mohan, and
  Ranka]{anand1993improved}
Anand, R., Mehrotra, K., Mohan, C., and Ranka, S.
\newblock An improved algorithm for neural network classification of imbalanced
  training sets.
\newblock \emph{IEEE Transactions on Neural Networks}, 4\penalty0 (6):\penalty0
  962--969, 1993.
\newblock \doi{10.1109/72.286891}.
\newblock URL \url{https://doi.org/10.1109/72.286891}.

\bibitem[Ansari \& White(2024)Ansari and White]{ansari2023peptide}
Ansari, M. and White, A.~D.
\newblock Learning peptide properties with positive examples only.
\newblock \emph{Digital Discovery}, 3:\penalty0 977--986, 2024.
\newblock \doi{10.1039/D3DD00218G}.
\newblock URL \url{http://dx.doi.org/10.1039/D3DD00218G}.

\bibitem[Baroffio et~al.(2024)Baroffio, Rotondo, and
  Gherardi]{baroffio2024resolution}
Baroffio, A., Rotondo, P., and Gherardi, M.
\newblock Resolution of similar patterns in a solvable model of unsupervised
  deep learning with structured data.
\newblock \emph{Chaos, Solitons \& Fractals}, 182:\penalty0 114848, 2024.
\newblock ISSN 0960-0779.
\newblock \doi{10.1016/j.chaos.2024.114848}.
\newblock URL
  \url{https://www.sciencedirect.com/science/article/pii/S0960077924004004}.

\bibitem[Batista et~al.(2004)Batista, Prati, and Monard]{batista2004methods}
Batista, G. E. A. P.~A., Prati, R.~C., and Monard, M.~C.
\newblock A study of the behavior of several methods for balancing machine
  learning training data.
\newblock \emph{SIGKDD Explor. Newsl.}, 6\penalty0 (1):\penalty0 20–29, jun
  2004.
\newblock ISSN 1931-0145.
\newblock \doi{10.1145/1007730.1007735}.
\newblock URL \url{https://doi.org/10.1145/1007730.1007735}.

\bibitem[Bordelon et~al.(2020)Bordelon, Canatar, and
  Pehlevan]{bordelon2020spectrum}
Bordelon, B., Canatar, A., and Pehlevan, C.
\newblock Spectrum dependent learning curves in kernel regression and wide
  neural networks.
\newblock In III, H.~D. and Singh, A. (eds.), \emph{Proceedings of the 37th
  International Conference on Machine Learning}, volume 119 of
  \emph{Proceedings of Machine Learning Research}, pp.\  1024--1034. PMLR,
  13--18 Jul 2020.
\newblock URL \url{https://proceedings.mlr.press/v119/bordelon20a.html}.

\bibitem[Buda et~al.(2018)Buda, Maki, and Mazurowski]{buda2018cnn}
Buda, M., Maki, A., and Mazurowski, M.~A.
\newblock A systematic study of the class imbalance problem in convolutional
  neural networks.
\newblock \emph{Neural Networks}, 106:\penalty0 249--259, 2018.
\newblock ISSN 0893-6080.
\newblock \doi{10.1016/j.neunet.2018.07.011}.
\newblock URL
  \url{https://www.sciencedirect.com/science/article/pii/S0893608018302107}.

\bibitem[Carbone et~al.(2023)Carbone, Decelle, Rosset, and
  Seoane]{carbone2023fast}
Carbone, A., Decelle, A., Rosset, L., and Seoane, B.
\newblock Fast and functional structured data generator.
\newblock In \emph{ICML 2023 Workshop on Structured Probabilistic Inference
  {\&} Generative Modeling}, 2023.
\newblock URL \url{https://openreview.net/forum?id=uXkfPvjYeM}.

\bibitem[Chaudhuri et~al.(2023)Chaudhuri, Ahuja, Arjovsky, and
  Lopez-Paz]{chaudhuri2023}
Chaudhuri, K., Ahuja, K., Arjovsky, M., and Lopez-Paz, D.
\newblock Why does throwing away data improve worst-group error?
\newblock In Krause, A., Brunskill, E., Cho, K., Engelhardt, B., Sabato, S.,
  and Scarlett, J. (eds.), \emph{Proceedings of the 40th International
  Conference on Machine Learning}, volume 202 of \emph{Proceedings of Machine
  Learning Research}, pp.\  4144--4188. PMLR, 23--29 Jul 2023.
\newblock URL \url{https://proceedings.mlr.press/v202/chaudhuri23a.html}.

\bibitem[Chawla et~al.(2002)Chawla, Bowyer, Hall, and
  Kegelmeyer]{chawla2002smote}
Chawla, N.~V., Bowyer, K.~W., Hall, L.~O., and Kegelmeyer, W.~P.
\newblock {SMOTE}: synthetic minority over-sampling technique.
\newblock \emph{Journal of artificial intelligence research}, 16:\penalty0
  321--357, 2002.
\newblock \doi{10.1613/jair.953}.
\newblock URL \url{https://doi.org/10.1613/jair.953}.

\bibitem[Cheng et~al.(2015)Cheng, Zhou, and Guan]{cheng2015proteinRNA}
Cheng, Z., Zhou, S., and Guan, J.
\newblock Computationally predicting protein-{RNA} interactions using only
  positive and unlabeled examples.
\newblock \emph{Journal of Bioinformatics and Computational Biology},
  13\penalty0 (03):\penalty0 1541005, 2015.
\newblock \doi{10.1142/S021972001541005X}.
\newblock URL \url{https://doi.org/10.1142/S021972001541005X}.

\bibitem[Chung et~al.(2018)Chung, Lee, and Sompolinsky]{chung2018manifold}
Chung, S., Lee, D.~D., and Sompolinsky, H.
\newblock Classification and geometry of general perceptual manifolds.
\newblock \emph{Phys. Rev. X}, 8:\penalty0 031003, Jul 2018.
\newblock \doi{10.1103/PhysRevX.8.031003}.
\newblock URL \url{https://link.aps.org/doi/10.1103/PhysRevX.8.031003}.

\bibitem[Cui et~al.(2023)Cui, Krzakala, and Zdeborova]{cui2023optimal}
Cui, H., Krzakala, F., and Zdeborova, L.
\newblock {B}ayes-optimal learning of deep random networks of extensive-width.
\newblock In Krause, A., Brunskill, E., Cho, K., Engelhardt, B., Sabato, S.,
  and Scarlett, J. (eds.), \emph{Proceedings of the 40th International
  Conference on Machine Learning}, volume 202 of \emph{Proceedings of Machine
  Learning Research}, pp.\  6468--6521. PMLR, 23--29 Jul 2023.
\newblock URL \url{https://proceedings.mlr.press/v202/cui23b.html}.

\bibitem[Cui et~al.(2024)Cui, Pesce, Dandi, Krzakala, Lu, Zdeborova, and
  Loureiro]{cui2024asymptotics}
Cui, H., Pesce, L., Dandi, Y., Krzakala, F., Lu, Y., Zdeborova, L., and
  Loureiro, B.
\newblock Asymptotics of feature learning in two-layer networks after one
  gradient-step.
\newblock In Salakhutdinov, R., Kolter, Z., Heller, K., Weller, A., Oliver, N.,
  Scarlett, J., and Berkenkamp, F. (eds.), \emph{Proceedings of the 41st
  International Conference on Machine Learning}, volume 235 of
  \emph{Proceedings of Machine Learning Research}, pp.\  9662--9695. PMLR,
  21--27 Jul 2024.
\newblock URL \url{https://proceedings.mlr.press/v235/cui24d.html}.

\bibitem[Dandi et~al.(2023)Dandi, Stephan, Krzakala, Loureiro, and
  Zdeborov\'{a}]{dandi2023universality}
Dandi, Y., Stephan, L., Krzakala, F., Loureiro, B., and Zdeborov\'{a}, L.
\newblock Universality laws for {G}aussian mixtures in generalized linear
  models.
\newblock In Oh, A., Naumann, T., Globerson, A., Saenko, K., Hardt, M., and
  Levine, S. (eds.), \emph{Advances in Neural Information Processing Systems},
  volume~36, pp.\  54754--54768. Curran Associates, Inc., 2023.
\newblock URL
  \url{https://proceedings.neurips.cc/paper_files/paper/2023/file/abccb8a90b30d45b948360ba41f5a20f-Paper-Conference.pdf}.

\bibitem[{Del Giudice, P.} et~al.(1989){Del Giudice, P.}, {Franz, S.}, and
  {Virasoro, M. A.}]{delgiudice1989}
{Del Giudice, P.}, {Franz, S.}, and {Virasoro, M. A.}
\newblock Perceptron beyond the limit of capacity.
\newblock \emph{J. Phys. France}, 50\penalty0 (2):\penalty0 121--134, 1989.
\newblock \doi{10.1051/jphys:01989005002012100}.
\newblock URL \url{https://doi.org/10.1051/jphys:01989005002012100}.

\bibitem[Dietrich et~al.(1999)Dietrich, Opper, and
  Sompolinsky]{dietrich1999svm}
Dietrich, R., Opper, M., and Sompolinsky, H.
\newblock Statistical mechanics of support vector networks.
\newblock \emph{Phys. Rev. Lett.}, 82:\penalty0 2975--2978, 04 1999.
\newblock \doi{10.1103/PhysRevLett.82.2975}.
\newblock URL \url{https://link.aps.org/doi/10.1103/PhysRevLett.82.2975}.

\bibitem[Douzas \& Bacao(2018)Douzas and Bacao]{douzas2018gan}
Douzas, G. and Bacao, F.
\newblock Effective data generation for imbalanced learning using conditional
  generative adversarial networks.
\newblock \emph{Expert Systems with Applications}, 91:\penalty0 464--471, 2018.
\newblock ISSN 0957-4174.
\newblock \doi{10.1016/j.eswa.2017.09.030}.
\newblock URL
  \url{https://www.sciencedirect.com/science/article/pii/S0957417417306346}.

\bibitem[Engel \& Van~den Broeck(2001)Engel and Van~den Broeck]{engel2001book}
Engel, A. and Van~den Broeck, C.
\newblock \emph{Statistical Mechanics of Learning}.
\newblock Cambridge University Press, 2001.
\newblock \doi{10.1017/CBO9781139164542}.
\newblock URL \url{https://doi.org/10.1017/CBO9781139164542}.

\bibitem[Fern{\'a}ndez et~al.(2018)Fern{\'a}ndez, Garc{\'i}a, Galar, Prati,
  Krawczyk, and Herrera]{fernandez2018book}
Fern{\'a}ndez, A., Garc{\'i}a, S., Galar, M., Prati, R.~C., Krawczyk, B., and
  Herrera, F.
\newblock \emph{Learning from Imbalanced Data Sets}.
\newblock Springer Cham, 2018.
\newblock \doi{10.1007/978-3-319-98074-4}.
\newblock URL \url{https://doi.org/10.1007/978-3-319-98074-4}.

\bibitem[Fotouhi et~al.(2019)Fotouhi, Asadi, and Kattan]{fotouhi2019cancer}
Fotouhi, S., Asadi, S., and Kattan, M.~W.
\newblock A comprehensive data level analysis for cancer diagnosis on
  imbalanced data.
\newblock \emph{Journal of Biomedical Informatics}, 90:\penalty0 103089, 2019.
\newblock ISSN 1532-0464.
\newblock \doi{10.1016/j.jbi.2018.12.003}.
\newblock URL
  \url{https://www.sciencedirect.com/science/article/pii/S1532046418302302}.

\bibitem[Francazi et~al.(2023)Francazi, Baity-Jesi, and
  Lucchi]{francazi2023theoretical}
Francazi, E., Baity-Jesi, M., and Lucchi, A.
\newblock A theoretical analysis of the learning dynamics under class
  imbalance.
\newblock In Krause, A., Brunskill, E., Cho, K., Engelhardt, B., Sabato, S.,
  and Scarlett, J. (eds.), \emph{Proceedings of the 40th International
  Conference on Machine Learning}, volume 202 of \emph{Proceedings of Machine
  Learning Research}, pp.\  10285--10322. PMLR, 23--29 Jul 2023.
\newblock URL \url{https://proceedings.mlr.press/v202/francazi23a.html}.

\bibitem[Franz et~al.(2017)Franz, Parisi, Sevelev, Urbani, and
  Zamponi]{urbani2017jamming}
Franz, S., Parisi, G., Sevelev, M., Urbani, P., and Zamponi, F.
\newblock {Universality of the SAT-UNSAT (jamming) threshold in non-convex
  continuous constraint satisfaction problems}.
\newblock \emph{SciPost Phys.}, 2:\penalty0 019, 2017.
\newblock \doi{10.21468/SciPostPhys.2.3.019}.
\newblock URL \url{https://scipost.org/10.21468/SciPostPhys.2.3.019}.

\bibitem[Franz et~al.(2019)Franz, Sclocchi, and Urbani]{franz2019perceptron}
Franz, S., Sclocchi, A., and Urbani, P.
\newblock Critical jammed phase of the linear perceptron.
\newblock \emph{Phys. Rev. Lett.}, 123:\penalty0 115702, Sep 2019.
\newblock \doi{10.1103/PhysRevLett.123.115702}.
\newblock URL \url{https://link.aps.org/doi/10.1103/PhysRevLett.123.115702}.

\bibitem[Gardner(1988)]{gardner1988space}
Gardner, E.
\newblock The space of interactions in neural network models.
\newblock \emph{Journal of Physics A: Mathematical and General}, 21\penalty0
  (1):\penalty0 257, Jan 1988.
\newblock \doi{10.1088/0305-4470/21/1/030}.
\newblock URL \url{https://dx.doi.org/10.1088/0305-4470/21/1/030}.

\bibitem[Gerace et~al.(2020)Gerace, Loureiro, Krzakala, Mezard, and
  Zdeborova]{gerace2020generalisation}
Gerace, F., Loureiro, B., Krzakala, F., Mezard, M., and Zdeborova, L.
\newblock Generalisation error in learning with random features and the hidden
  manifold model.
\newblock In III, H.~D. and Singh, A. (eds.), \emph{Proceedings of the 37th
  International Conference on Machine Learning}, volume 119 of
  \emph{Proceedings of Machine Learning Research}, pp.\  3452--3462. PMLR,
  13--18 Jul 2020.
\newblock URL \url{https://proceedings.mlr.press/v119/gerace20a.html}.

\bibitem[Ghosh et~al.(2022)Ghosh, Bellinger, Corizzo, Branco, Krawczyk, and
  Japkowicz]{ghosh2022deep}
Ghosh, K., Bellinger, C., Corizzo, R., Branco, P., Krawczyk, B., and Japkowicz,
  N.
\newblock The class imbalance problem in deep learning.
\newblock \emph{Machine Learning}, Dec 2022.
\newblock ISSN 1573-0565.
\newblock URL \url{https://doi.org/10.1007/s10994-022-06268-8}.

\bibitem[Goldt et~al.(2020)Goldt, M\'ezard, Krzakala, and
  Zdeborov\'a]{goldt2020gep}
Goldt, S., M\'ezard, M., Krzakala, F., and Zdeborov\'a, L.
\newblock Modeling the influence of data structure on learning in neural
  networks: The hidden manifold model.
\newblock \emph{Phys. Rev. X}, 10:\penalty0 041044, Dec 2020.
\newblock \doi{10.1103/PhysRevX.10.041044}.
\newblock URL \url{https://link.aps.org/doi/10.1103/PhysRevX.10.041044}.

\bibitem[He \& Ma(2013)He and Ma]{he2013book}
He, H. and Ma, Y. (eds.).
\newblock \emph{Imbalanced Learning: Foundations, Algorithms, and
  Applications}.
\newblock {John Wiley \& Sons, Ltd}, 2013.
\newblock ISBN 9781118646106.
\newblock \doi{10.1002/9781118646106}.
\newblock URL
  \url{https://onlinelibrary.wiley.com/doi/abs/10.1002/9781118646106}.

\bibitem[He et~al.(2008)He, Bai, Garcia, and Li]{he2008adasyn}
He, H., Bai, Y., Garcia, E.~A., and Li, S.
\newblock {ADASYN}: Adaptive synthetic sampling approach for imbalanced
  learning.
\newblock In \emph{2008 IEEE International Joint Conference on Neural Networks
  (IEEE World Congress on Computational Intelligence)}, pp.\  1322--1328, 2008.
\newblock \doi{10.1109/IJCNN.2008.4633969}.
\newblock URL \url{https://doi.org/10.1109/IJCNN.2008.4633969}.

\bibitem[Jacot et~al.(2018)Jacot, Gabriel, and Hongler]{jacot2018NTK}
Jacot, A., Gabriel, F., and Hongler, C.
\newblock Neural tangent kernel: Convergence and generalization in neural
  networks.
\newblock In Bengio, S., Wallach, H., Larochelle, H., Grauman, K.,
  Cesa-Bianchi, N., and Garnett, R. (eds.), \emph{Advances in Neural
  Information Processing Systems}, volume~31. Curran Associates, Inc., 2018.
\newblock URL
  \url{https://proceedings.neurips.cc/paper/2018/file/5a4be1fa34e62bb8a6ec6b91d2462f5a-Paper.pdf}.

\bibitem[Japkowicz \& Stephen(2002)Japkowicz and
  Stephen]{japkowicz2002systematic}
Japkowicz, N. and Stephen, S.
\newblock The class imbalance problem: A systematic study.
\newblock \emph{Intelligent Data Analysis}, 6:\penalty0 429--449, 2002.
\newblock ISSN 1571-4128.
\newblock \doi{10.3233/IDA-2002-6504}.
\newblock URL \url{https://doi.org/10.3233/IDA-2002-6504}.

\bibitem[Johnson \& Khoshgoftaar(2019)Johnson and
  Khoshgoftaar]{johnson2019deep}
Johnson, J.~M. and Khoshgoftaar, T.~M.
\newblock Survey on deep learning with class imbalance.
\newblock \emph{Journal of Big Data}, 6\penalty0 (1):\penalty0 27, Mar 2019.
\newblock ISSN 2196-1115.
\newblock \doi{10.1186/s40537-019-0192-5}.
\newblock URL \url{https://doi.org/10.1186/s40537-019-0192-5}.

\bibitem[Jones et~al.(2001)Jones, Oliphant, and Peterson]{jones_scipy:_2001}
Jones, E., Oliphant, T., and Peterson, P.
\newblock {SciPy:} open source scientific tools for {Python}, 2001.
\newblock URL \url{http://www.scipy.org}.

\bibitem[Jurafsky \& Martin(2009)Jurafsky and Martin]{jurafsky2009speech}
Jurafsky, D. and Martin, J.
\newblock \emph{Speech and Language Processing: An Introduction to Natural
  Language Processing, Computational Linguistics, and Speech Recognition}.
\newblock Prentice Hall series in artificial intelligence. Pearson Prentice
  Hall, 2009.
\newblock ISBN 9780131873216.
\newblock URL \url{https://books.google.fr/books?id=fZmj5UNK8AQC}.

\bibitem[Kepler \& Abbott(1988)Kepler and Abbott]{abbott1989domains}
Kepler, T.~B. and Abbott, L.
\newblock Domains of attraction in neural networks.
\newblock \emph{J. Phys. France}, 49\penalty0 (10):\penalty0 1657--1662, 1988.
\newblock \doi{10.1051/jphys:0198800490100165700}.
\newblock URL \url{https://doi.org/10.1051/jphys:0198800490100165700}.

\bibitem[Kovács(2019)]{kovacs2019emp}
Kovács, G.
\newblock An empirical comparison and evaluation of minority oversampling
  techniques on a large number of imbalanced datasets.
\newblock \emph{Applied Soft Computing}, 83:\penalty0 105662, 2019.
\newblock ISSN 1568-4946.
\newblock \doi{10.1016/j.asoc.2019.105662}.
\newblock URL
  \url{https://www.sciencedirect.com/science/article/pii/S1568494619304429}.

\bibitem[Krawczyk et~al.(2016)Krawczyk, Galar, Łukasz Jeleń, and
  Herrera]{krawczyk2016cancer}
Krawczyk, B., Galar, M., Łukasz Jeleń, and Herrera, F.
\newblock Evolutionary undersampling boosting for imbalanced classification of
  breast cancer malignancy.
\newblock \emph{Applied Soft Computing}, 38:\penalty0 714--726, 2016.
\newblock ISSN 1568-4946.
\newblock \doi{10.1016/j.asoc.2015.08.060}.
\newblock URL
  \url{https://www.sciencedirect.com/science/article/pii/S1568494615005815}.

\bibitem[Kubat \& Matwin(1997)Kubat and Matwin]{kubat1997curse}
Kubat, M. and Matwin, S.
\newblock Addressing the curse of imbalanced training sets: One-sided
  selection.
\newblock In Fisher, D.~H. (ed.), \emph{Proceedings of the Fourteenth
  International Conference on Machine Learning {(ICML} 1997), Nashville,
  Tennessee, USA, July 8-12, 1997}, pp.\  179--186. Morgan Kaufmann, 1997.

\bibitem[Laurikkala(2001)]{laurikkala2001improving}
Laurikkala, J.
\newblock Improving identification of difficult small classes by balancing
  class distribution.
\newblock In Quaglini, S., Barahona, P., and Andreassen, S. (eds.),
  \emph{Artificial Intelligence in Medicine}, pp.\  63--66, Berlin, Heidelberg,
  2001. Springer Berlin Heidelberg.
\newblock ISBN 978-3-540-48229-1.
\newblock URL \url{https://doi.org/10.1007/3-540-48229-6_9}.

\bibitem[Lee et~al.(2018)Lee, Sohl-dickstein, Pennington, Novak, Schoenholz,
  and Bahri]{lee2018gaussian}
Lee, J., Sohl-dickstein, J., Pennington, J., Novak, R., Schoenholz, S., and
  Bahri, Y.
\newblock Deep neural networks as {G}aussian processes.
\newblock In \emph{International Conference on Learning Representations}, 2018.
\newblock URL \url{https://openreview.net/forum?id=B1EA-M-0Z}.

\bibitem[Lema{{\^i}}tre et~al.(2017)Lema{{\^i}}tre, Nogueira, and
  Aridas]{lemaitre2017curse}
Lema{{\^i}}tre, G., Nogueira, F., and Aridas, C.~K.
\newblock Imbalanced-learn: A {P}ython toolbox to tackle the curse of
  imbalanced datasets in machine learning.
\newblock \emph{Journal of Machine Learning Research}, 18\penalty0
  (17):\penalty0 1--5, 2017.
\newblock URL \url{http://jmlr.org/papers/v18/16-365.html}.

\bibitem[Lemnaru \& Potolea(2012)Lemnaru and Potolea]{lemnaru2012systematic}
Lemnaru, C. and Potolea, R.
\newblock Imbalanced classification problems: Systematic study, issues and best
  practices.
\newblock In Zhang, R., Zhang, J., Zhang, Z., Filipe, J., and Cordeiro, J.
  (eds.), \emph{Enterprise Information Systems}, pp.\  35--50, Berlin,
  Heidelberg, 2012. Springer Berlin Heidelberg.
\newblock ISBN 978-3-642-29958-2.
\newblock \doi{10.1007/978-3-642-29958-2_3}.
\newblock URL \url{https://doi.org/10.1007/978-3-642-29958-2_3}.

\bibitem[Li et~al.(1996)Li, Helling, Tang, and Wingreen]{li1996emergence}
Li, H., Helling, R., Tang, C., and Wingreen, N.
\newblock Emergence of preferred structures in a simple model of protein
  folding.
\newblock \emph{Science}, 273\penalty0 (5275):\penalty0 666--669, 1996.
\newblock \doi{10.1126/science.273.5275.666}.
\newblock URL
  \url{https://www.science.org/doi/abs/10.1126/science.273.5275.666}.

\bibitem[Liu et~al.(2017)Liu, Wang, Zhang, Chen, and Xiang]{liu2017twitter}
Liu, S., Wang, Y., Zhang, J., Chen, C., and Xiang, Y.
\newblock Addressing the class imbalance problem in twitter spam detection
  using ensemble learning.
\newblock \emph{{Computers \& Security}}, 69:\penalty0 35--49, 2017.
\newblock ISSN 0167-4048.
\newblock \doi{10.1016/j.cose.2016.12.004}.
\newblock URL
  \url{https://www.sciencedirect.com/science/article/pii/S0167404816301754}.

\bibitem[Liu et~al.(2009)Liu, Loh, and Sun]{liu2009text}
Liu, Y., Loh, H.~T., and Sun, A.
\newblock Imbalanced text classification: A term weighting approach.
\newblock \emph{Expert Systems with Applications}, 36\penalty0 (1):\penalty0
  690--701, 2009.
\newblock ISSN 0957-4174.
\newblock \doi{10.1016/j.eswa.2007.10.042}.
\newblock URL
  \url{https://www.sciencedirect.com/science/article/pii/S0957417407005350}.

\bibitem[Lopez et~al.(1995)Lopez, Schroder, and Opper]{lopez1995storage}
Lopez, B., Schroder, M., and Opper, M.
\newblock Storage of correlated patterns in a perceptron.
\newblock \emph{Journal of Physics A: Mathematical and General}, 28\penalty0
  (16):\penalty0 L447, aug 1995.
\newblock \doi{10.1088/0305-4470/28/16/005}.
\newblock URL \url{https://dx.doi.org/10.1088/0305-4470/28/16/005}.

\bibitem[Loureiro et~al.(2021)Loureiro, Sicuro, Gerbelot, Pacco, Krzakala, and
  Zdeborov\'{a}]{loureiro2021gmm}
Loureiro, B., Sicuro, G., Gerbelot, C., Pacco, A., Krzakala, F., and
  Zdeborov\'{a}, L.
\newblock Learning gaussian mixtures with generalized linear models: Precise
  asymptotics in high-dimensions.
\newblock In Ranzato, M., Beygelzimer, A., Dauphin, Y., Liang, P., and Vaughan,
  J.~W. (eds.), \emph{Advances in Neural Information Processing Systems},
  volume~34, pp.\  10144--10157. Curran Associates, Inc., 2021.
\newblock URL
  \url{https://proceedings.neurips.cc/paper_files/paper/2021/file/543e83748234f7cbab21aa0ade66565f-Paper.pdf}.

\bibitem[Mannelli et~al.(2024)Mannelli, Gerace, Rostamzadeh, and
  Saglietti]{mannelli2023unfair}
Mannelli, S.~S., Gerace, F., Rostamzadeh, N., and Saglietti, L.
\newblock Bias-inducing geometries: exactly solvable data model with fairness
  implications.
\newblock In \emph{ICML 2024 Workshop on Geometry-grounded Representation
  Learning and Generative Modeling}, 2024.
\newblock URL \url{https://openreview.net/forum?id=oupizzpMpY}.

\bibitem[Menon et~al.(2013)Menon, Narasimhan, Agarwal, and
  Chawla]{menon2013statistical}
Menon, A., Narasimhan, H., Agarwal, S., and Chawla, S.
\newblock On the statistical consistency of algorithms for binary
  classification under class imbalance.
\newblock In Dasgupta, S. and McAllester, D. (eds.), \emph{Proceedings of the
  30th International Conference on Machine Learning}, volume 28-3 of
  \emph{Proceedings of Machine Learning Research}, pp.\  603--611, Atlanta,
  Georgia, USA, 17--19 Jun 2013. PMLR.
\newblock URL \url{https://proceedings.mlr.press/v28/menon13a.html}.

\bibitem[Mezard(1989)]{mezard1989cavity}
Mezard, M.
\newblock The space of interactions in neural networks: Gardner's computation
  with the cavity method.
\newblock \emph{Journal of Physics A: Mathematical and General}, 22\penalty0
  (12):\penalty0 2181, jun 1989.
\newblock \doi{10.1088/0305-4470/22/12/018}.
\newblock URL \url{https://dx.doi.org/10.1088/0305-4470/22/12/018}.

\bibitem[Mignacco et~al.(2020)Mignacco, Krzakala, Lu, Urbani, and
  Zdeborova]{mignacco2020gmm}
Mignacco, F., Krzakala, F., Lu, Y., Urbani, P., and Zdeborova, L.
\newblock The role of regularization in classification of high-dimensional
  noisy {G}aussian mixture.
\newblock In III, H.~D. and Singh, A. (eds.), \emph{Proceedings of the 37th
  International Conference on Machine Learning}, volume 119 of
  \emph{Proceedings of Machine Learning Research}, pp.\  6874--6883. PMLR,
  13--18 Jul 2020.
\newblock URL \url{https://proceedings.mlr.press/v119/mignacco20a.html}.

\bibitem[Mirny \& Shakhnovich(2001)Mirny and Shakhnovich]{mirny2001protein}
Mirny, L. and Shakhnovich, E.
\newblock Protein folding theory: From lattice to all-atom models.
\newblock \emph{Annual Review of Biophysics}, 30\penalty0 (Volume 30,
  2001):\penalty0 361--396, 2001.
\newblock ISSN 1936-1238.
\newblock \doi{10.1146/annurev.biophys.30.1.361}.
\newblock URL
  \url{https://www.annualreviews.org/content/journals/10.1146/annurev.biophys.30.1.361}.

\bibitem[Mirza et~al.(2021)Mirza, Haroon, Khan, Padhani, and
  Syed]{mirza2021deep}
Mirza, B., Haroon, D., Khan, B., Padhani, A., and Syed, T.~Q.
\newblock Deep generative models to counter class imbalance: A model-metric
  mapping with proportion calibration methodology.
\newblock \emph{IEEE Access}, 9:\penalty0 55879--55897, 2021.
\newblock \doi{10.1109/ACCESS.2021.3071389}.
\newblock URL \url{https://doi.org/10.1109/ACCESS.2021.3071389}.

\bibitem[Monasson(1992)]{monasson1992}
Monasson, R.
\newblock Properties of neural networks storing spatially correlated patterns.
\newblock \emph{Journal of Physics A: Mathematical and General}, 25\penalty0
  (13):\penalty0 3701, jul 1992.
\newblock \doi{10.1088/0305-4470/25/13/019}.
\newblock URL \url{https://dx.doi.org/10.1088/0305-4470/25/13/019}.

\bibitem[Mondal et~al.(2023)Mondal, Singhal, Tiwary, Singla, and
  {AP}]{mondal2023ae}
Mondal, A.~K., Singhal, L., Tiwary, P., Singla, P., and {AP}, P.
\newblock Minority oversampling for imbalanced data via class-preserving
  regularized auto-encoders.
\newblock In Ruiz, F., Dy, J., and van~de Meent, J.-W. (eds.),
  \emph{Proceedings of The 26th International Conference on Artificial
  Intelligence and Statistics}, volume 206 of \emph{Proceedings of Machine
  Learning Research}, pp.\  3440--3465. PMLR, 25--27 Apr 2023.
\newblock URL \url{https://proceedings.mlr.press/v206/mondal23a.html}.

\bibitem[Nadal \& Rau(1991)Nadal and Rau]{nadal1991potts}
Nadal, J.-P. and Rau, A.
\newblock Storage capacity of a potts-perceptron.
\newblock \emph{J. Phys. I France}, 1\penalty0 (8):\penalty0 1109--1121, 1991.
\newblock \doi{10.1051/jp1:1991104}.
\newblock URL \url{https://doi.org/10.1051/jp1:1991104}.

\bibitem[Naveh \& Ringel(2021)Naveh and Ringel]{naveh2021}
Naveh, G. and Ringel, Z.
\newblock A self consistent theory of gaussian processes captures feature
  learning effects in finite {CNNs}.
\newblock In Ranzato, M., Beygelzimer, A., Dauphin, Y., Liang, P., and Vaughan,
  J.~W. (eds.), \emph{Advances in Neural Information Processing Systems},
  volume~34, pp.\  21352--21364. Curran Associates, Inc., 2021.
\newblock URL
  \url{https://proceedings.neurips.cc/paper_files/paper/2021/file/b24d21019de5e59da180f1661904f49a-Paper.pdf}.

\bibitem[Pacelli et~al.(2023)Pacelli, Ariosto, Pastore, Ginelli, Gherardi, and
  Rotondo]{pacelli2022finitewidth}
Pacelli, R., Ariosto, S., Pastore, M., Ginelli, F., Gherardi, M., and Rotondo,
  P.
\newblock A statistical mechanics framework for bayesian deep neural networks
  beyond the infinite-width limit.
\newblock \emph{Nature Machine Intelligence}, 5\penalty0 (12):\penalty0
  1497--1507, Dec 2023.
\newblock ISSN 2522-5839.
\newblock \doi{10.1038/s42256-023-00767-6}.
\newblock URL \url{https://doi.org/10.1038/s42256-023-00767-6}.

\bibitem[Pastore(2021)]{pastore2021critical}
Pastore, M.
\newblock Critical properties of the sat/unsat transitions in the
  classification problem of structured data.
\newblock \emph{Journal of Statistical Mechanics: Theory and Experiment},
  2021\penalty0 (11):\penalty0 113301, 11 2021.
\newblock \doi{10.1088/1742-5468/ac312b}.
\newblock URL \url{https://dx.doi.org/10.1088/1742-5468/ac312b}.

\bibitem[Pastore et~al.(2020)Pastore, Rotondo, Erba, and
  Gherardi]{pastore2020structure}
Pastore, M., Rotondo, P., Erba, V., and Gherardi, M.
\newblock Statistical learning theory of structured data.
\newblock \emph{Phys. Rev. E}, 102:\penalty0 032119, Sep 2020.
\newblock \doi{10.1103/PhysRevE.102.032119}.
\newblock URL \url{https://link.aps.org/doi/10.1103/PhysRevE.102.032119}.

\bibitem[Pedregosa et~al.(2011)Pedregosa, Varoquaux, Gramfort, Michel, Thirion,
  Grisel, Blondel, Prettenhofer, Weiss, Dubourg, Vanderplas, Passos,
  Cournapeau, Brucher, Perrot, and Duchesnay]{pedregosa2011scikitlearn}
Pedregosa, F., Varoquaux, G., Gramfort, A., Michel, V., Thirion, B., Grisel,
  O., Blondel, M., Prettenhofer, P., Weiss, R., Dubourg, V., Vanderplas, J.,
  Passos, A., Cournapeau, D., Brucher, M., Perrot, M., and Duchesnay, E.
\newblock Scikit-learn: Machine learning in {P}ython.
\newblock \emph{Journal of Machine Learning Research}, 12:\penalty0 2825--2830,
  2011.
\newblock URL
  \url{http://www.jmlr.org/papers/volume12/pedregosa11a/pedregosa11a.pdf}.

\bibitem[Pesce et~al.(2023)Pesce, Krzakala, Loureiro, and Stephan]{pesce2023}
Pesce, L., Krzakala, F., Loureiro, B., and Stephan, L.
\newblock Are {G}aussian data all you need? {T}he extents and limits of
  universality in high-dimensional generalized linear estimation.
\newblock In Krause, A., Brunskill, E., Cho, K., Engelhardt, B., Sabato, S.,
  and Scarlett, J. (eds.), \emph{Proceedings of the 40th International
  Conference on Machine Learning}, volume 202 of \emph{Proceedings of Machine
  Learning Research}, pp.\  27680--27708. PMLR, 23--29 Jul 2023.
\newblock URL \url{https://proceedings.mlr.press/v202/pesce23a.html}.

\bibitem[Rotondo et~al.(2020)Rotondo, Pastore, and Gherardi]{rotondo2020beyond}
Rotondo, P., Pastore, M., and Gherardi, M.
\newblock Beyond the storage capacity: Data-driven satisfiability transition.
\newblock \emph{Phys. Rev. Lett.}, 125:\penalty0 120601, Sep 2020.
\newblock \doi{10.1103/PhysRevLett.125.120601}.
\newblock URL \url{https://link.aps.org/doi/10.1103/PhysRevLett.125.120601}.

\bibitem[Sauvola \& Pietikäinen(2000)Sauvola and Pietikäinen]{SAUVOLA2000225}
Sauvola, J. and Pietikäinen, M.
\newblock Adaptive document image binarization.
\newblock \emph{Pattern Recognition}, 33\penalty0 (2):\penalty0 225--236, 2000.
\newblock ISSN 0031-3203.
\newblock \doi{10.1016/S0031-3203(99)00055-2}.
\newblock URL
  \url{https://www.sciencedirect.com/science/article/pii/S0031320399000552}.

\bibitem[Shakhnovich \& Gutin(1990)Shakhnovich and
  Gutin]{shakhnovich1990enumeration}
Shakhnovich, E. and Gutin, A.
\newblock Enumeration of all compact conformations of copolymers with random
  sequence of links.
\newblock \emph{The Journal of Chemical Physics}, 93\penalty0 (8):\penalty0
  5967--5971, 10 1990.
\newblock ISSN 0021-9606.
\newblock \doi{10.1063/1.459480}.
\newblock URL \url{https://doi.org/10.1063/1.459480}.

\bibitem[Shang et~al.(2023)Shang, Langlois, Tsioutsiouliklis, and
  Kang]{shang2023yahoo}
Shang, H., Langlois, J.-M., Tsioutsiouliklis, K., and Kang, C.
\newblock Precision/recall on imbalanced test data.
\newblock In Ruiz, F., Dy, J., and van~de Meent, J.-W. (eds.),
  \emph{Proceedings of The 26th International Conference on Artificial
  Intelligence and Statistics}, volume 206 of \emph{Proceedings of Machine
  Learning Research}, pp.\  9879--9891. PMLR, 25--27 Apr 2023.
\newblock URL \url{https://proceedings.mlr.press/v206/shang23a.html}.

\bibitem[Song et~al.(2021)Song, Bremer, Hinds, Raskutti, and
  Romero]{song2021proteinFunc}
Song, H., Bremer, B.~J., Hinds, E.~C., Raskutti, G., and Romero, P.~A.
\newblock Inferring protein sequence-function relationships with large-scale
  positive-unlabeled learning.
\newblock \emph{Cell Systems}, 12\penalty0 (1):\penalty0 92--101.e8, 2021.
\newblock ISSN 2405-4712.
\newblock \doi{10.1016/j.cels.2020.10.007}.
\newblock URL
  \url{https://www.sciencedirect.com/science/article/pii/S2405471220304142}.

\bibitem[Tieleman(2008)]{tieleman2008training}
Tieleman, T.
\newblock Training restricted {B}oltzmann machines using approximations to the
  likelihood gradient.
\newblock In \emph{Proceedings of the 25th International Conference on Machine
  Learning}, ICML '08, pp.\  1064–1071, New York, NY, USA, 2008. Association
  for Computing Machinery.
\newblock ISBN 9781605582054.
\newblock \doi{10.1145/1390156.1390290}.
\newblock URL \url{https://doi.org/10.1145/1390156.1390290}.

\bibitem[Wan et~al.(2017)Wan, Zhang, and He]{wan2017vae}
Wan, Z., Zhang, Y., and He, H.
\newblock Variational autoencoder based synthetic data generation for
  imbalanced learning.
\newblock In \emph{2017 IEEE Symposium Series on Computational Intelligence
  (SSCI)}, pp.\  1--7, 2017.
\newblock \doi{10.1109/SSCI.2017.8285168}.
\newblock URL \url{https://doi.org/10.1109/SSCI.2017.8285168}.

\bibitem[Wang et~al.(2006)Wang, Ding, Meraz, and Holbrook]{wang2006ncRNA}
Wang, C., Ding, C., Meraz, R.~F., and Holbrook, S.~R.
\newblock {PSoL: a positive sample only learning algorithm for finding
  non-coding RNA genes}.
\newblock \emph{Bioinformatics}, 22\penalty0 (21):\penalty0 2590--2596, 08
  2006.
\newblock ISSN 1367-4803.
\newblock \doi{10.1093/bioinformatics/btl441}.
\newblock URL \url{https://doi.org/10.1093/bioinformatics/btl441}.

\bibitem[Weiss(2004)]{weiss2004rarity}
Weiss, G.~M.
\newblock Mining with rarity: A unifying framework.
\newblock \emph{SIGKDD Explor. Newsl.}, 6\penalty0 (1):\penalty0 7–19, jun
  2004.
\newblock ISSN 1931-0145.
\newblock \doi{10.1145/1007730.1007734}.
\newblock URL \url{https://doi.org/10.1145/1007730.1007734}.

\bibitem[Wu et~al.(2020)Wu, Xu, Dai, Wan, Zhang, Yan, Tomizuka, Gonzalez,
  Keutzer, and Vajda]{wu2020visual}
Wu, B., Xu, C., Dai, X., Wan, A., Zhang, P., Yan, Z., Tomizuka, M., Gonzalez,
  J., Keutzer, K., and Vajda, P.
\newblock Visual transformers: Token-based image representation and processing
  for computer vision, 2020.
\newblock URL \url{https://arxiv.org/abs/2006.03677}.

\bibitem[Yang et~al.(2012)Yang, Li, Mei, Kwoh, and Ng]{yang2012gene}
Yang, P., Li, X.-L., Mei, J.-P., Kwoh, C.-K., and Ng, S.-K.
\newblock {Positive-unlabeled learning for disease gene identification}.
\newblock \emph{Bioinformatics}, 28\penalty0 (20):\penalty0 2640--2647, 08
  2012.
\newblock ISSN 1367-4803.
\newblock \doi{10.1093/bioinformatics/bts504}.
\newblock URL \url{https://doi.org/10.1093/bioinformatics/bts504}.

\bibitem[Zi{\k{e}}ba et~al.(2015)Zi{\k{e}}ba, Tomczak, and
  Gonczarek]{zieba2015rbm}
Zi{\k{e}}ba, M., Tomczak, J.~M., and Gonczarek, A.
\newblock {RBM-SMOTE}: {R}estricted {B}oltzmann machines for synthetic minority
  oversampling technique.
\newblock In Nguyen, N.~T., Trawi{\'{n}}ski, B., and Kosala, R. (eds.),
  \emph{Intelligent Information and Database Systems}, pp.\  377--386, Cham,
  2015. Springer International Publishing.
\newblock ISBN 978-3-319-15702-3.
\newblock URL \url{https://doi.org/10.1007/978-3-319-15702-3_37}.

\end{thebibliography}
\bibliographystyle{arxiv2024}

\newpage
\appendix
\onecolumn

\section{Details on the replica calculation\label{app:replicas}}

We give here all the details in order to get the theoretical predictions sketched in Sec.~\ref{sec:theo1} of the main text. The measure over the weights for the two cases of the spherical perceptron and the SVM in Eq.~\eqref{eq:Omega} is defined as
\begin{equation}
    \diff \mu (\vb{J})
    = \begin{cases}
    \prod_{i,t} \dfrac{\diff J_i(t)}{\Omega_0} \delta [\sum_u  J_i(u)]\delta ( \normLtwo{J} - L Q)  &\text{(perceptron)}\,,\\[2ex]
    \prod_{i,t} \dfrac{\diff J_i(t)}{\Omega_0} \delta[\sum_u  J_i(u)]\exp\!\left(- \dfrac{\beta \lambda}{2}\normLtwo{J} \right)  &\text{(SVM)}\,,
    \end{cases}
\label{eq:theoryJmeasure}
\end{equation}
where $\Omega_0$ is the corresponding normalization factor and for convenience we included the L2 regularization in the measure of the SVM. The common delta functions are enforcing the zero-sum conditions $\sum_t J_i(t) = 0$ for all $i$ mentioned in Sec.~\ref{sec:theo1}. We report in the following the case of the spherical perceptron, mentioning where the calculation differs crucially from the case of the SVM in Sec.~\ref{app:SVM}.

\subsection{Replicated partition function}
To evaluate the expected value of the logarithm of the partition function appearing in Eq.~\eqref{eq:logOmega}, we resort to the so-called replica trick from statistical physics: writing $\log \Omega = \lim_{n\to 0^+} (\Omega^n - 1)/n$, we convert the quenched averaged over the training data into an annealed average of the $n$-times replicated partition function $\Omega^n$. 
This quantity can be written as
\begin{equation}
    \begin{aligned}
        \mathbb{E}_\text{data} [\Omega^{n}]  =  \int \prod_{a=1}^n \diff b^a  \diff \mu(\vb{J}^a) & \prod_{y\in\{ \pm \}} \int \prod_{\mu=1}^{\alpha^y \! L}    \dfrac{\diff \Delta_{\mu}^a \diff \hat{\Delta}_{\mu}^a}{2 \pi} e^{-\beta (\kappa- y \Delta_{\mu}^a) \theta(\kappa -y \Delta_{\mu}^a ) } e^{ i \sum_{a, \mu} \hat{\Delta}_{\mu} ^a (\Delta_{\mu}^a +b)}  \mean{e^{ -i \sum_{a,\mu} \hat{\Delta}_{\mu}^a \frac{\vb{J}^a \cdot \vb{s}^\mu }{\sqrt{L}}  }}_{y} 
    \end{aligned}
    \label{eq:OmegaReplicated}
\end{equation}
where we inserted delta-functions for the variables (replicated versions of the ones defined in Eq.~\eqref{eq:Deltas})
\begin{equation}
    \Delta_{\mu} ^a = \dfrac{\vb{J}^a \cdot \vb{s}^{\mu}}{\sqrt{L}} -b\,,
\end{equation}
using their Fourier-conjugates $\hat{\Delta}^a_\mu$. We introduce the order parameters
\begin{equation}
\begin{aligned}
    r^a &=  \sum_{i,t} \dfrac{ J_i^a(t) \delta_i(t)}{L} \,,  \qquad
    q^{ab} &=  \sum_{i,j,t,u} \dfrac{J_i^a (t) J_j^b(u)}{L} C_{ij}(t,u),
\end{aligned}
\label{eq:orderParam}
\end{equation}

so that the average over the samples gives, for large $L$,
\begin{equation} 
    \begin{aligned}
    \mean{e^{ -i \sum_{a,\mu} \hat{\Delta}_{\mu}^a \frac{\vb{J}^a \cdot \vb{s}^\mu }{\sqrt{L}}  }}_{\pm} & \sim 
    \exp  \!\Big[ -\dfrac{1}{2} \sum_{a,b}  \hat{\Delta}_{\mu}^a \hat{\Delta}_{\mu}^b q^{ab} \mp \dfrac{i}{2} \sum_{a} \hat{\Delta}_{\mu}^a r^a  -\dfrac{i}{\sqrt{L}}\sum_a \hat{\Delta}_{\mu}^a \sum_{i,t} J_i^a(t) M_i(t) \Big]\,.
    \end{aligned}
\end{equation}
This asymptotic expansions is justified as long as higher moments in the distribution of the data can be neglected with respect to the first and second moments. The full integral to work out becomes
\begin{equation}
    \begin{aligned}
        &\mathbb{E}_\text{data}[\Omega^{n}]  = \int \prod_{a} \diff b^a \prod_{i ,t} \frac{\diff J_i^a(t)}{\Omega_0} \delta\Bigl[\sum_u J^a_i(u)\Bigr] \delta ( \lVert \vb{J}^a \rVert^2_2 - L Q ) \int \prod_{a} \diff r^{a} \delta ( \vb{J}^a \cdot \bm{\delta} - L r^{a} ) \\
        &\quad\times \int \prod_{a\le b} \diff q^{ab} \delta \Big( \sum_{i,j,t,u} J^a_i(t) J^b_i(u) C_{ij}(t,u) -L q^{ab} \Big)\\
        &\quad\times \prod_{y\in \{\pm\} } \int \prod_{a,\mu}    \dfrac{\diff \Delta_{\mu}^a \diff \hat{\Delta}_{\mu}^a}{2 \pi} e^{-\beta (\kappa-y\Delta_{\mu}^a) \theta(\kappa -y\Delta_{\mu}^a )  -\frac{1}{2} \sum_{\mu,a,b} \hat{\Delta}_{\mu}^a \hat{\Delta}_{\mu}^b q^{ab} +i \sum_{\mu,a}  \hat{\Delta}_{\mu} ^a \left[ \Delta_{\mu} ^a +b^a  - y \frac{r^a}{2} -  \frac{1}{\sqrt{L}} \sum_{i,t} J_i^a(t) M_i(t) \right] } \\
    \end{aligned}
    \label{eq:appOmegaReplicas}
\end{equation}

The last line depends explicitly on the weights $\vb{J}^a$ only through the combination $\vb{J}^a \cdot \vb{M}/\sqrt{L}$; however, as the threshold $b^a$ is a parameter to be optimized, it is always possible to shift it with
\begin{equation}
    b^a - \frac{1}{\sqrt{L}} \sum_{i,t} J_i^a(t) M_i(t) \to b^a \,,
\end{equation}
in order to absorb this term.

To perform the integrals over the weights, we express all the delta-functions in their Fourier representation, introducing additional variables $g_i^a$ (for the zero-sum conditions), $\hat{k}^a$ (for the spherical constraint) and $\hat{q}^{ab}$, $\hat{r}^a$ for the deltas enforcing the definition of the order parameters. Moreover, we impose the replica symmetric (RS) Ansatz
\begin{equation}
    b^a = b\,,\quad r^a = r\,,\quad \hat{k}^a = \hat{k}\,,\quad \hat{r}^a = \hat{r}\,, \quad q^{ab} = q_0\, \delta_{ab} + q(1-\delta_{ab})\, , \quad \hat{q}^{ab} = \hat{q}_0\, \delta_{ab} + \hat{q}(1-\delta_{ab})
    \label{eq:theoryRS}
\end{equation}
on all the tensors with replica indices. This Ansatz is justified as long as the optimization problem in Eq.~\eqref{eq:minimization-Jb} is convex; though one may find a non-convex regime arising for some values of $\alpha^\pm$, $\bm{\delta}$ and $C$, as discussed in~\citet{franz2019perceptron}, we expect it to be exact for many choices of the parameters. In practice, we find excellent agreement between our theoretical predictions and  numerical experiments throughout this work (see Fig.~\ref{fig:app_rs_theorysims}).

Under the Ansatz \eqref{eq:theoryRS}, the high-dimensional integral over weights and threshold can be carried out and reduces to a four-dimensional integral over the order parameters $b,r,q,q_0$. The calculation can be split in an entropic contribution, from the integrals over the weights, and an energetic term, from the training data. At the end, the expected value in Eq.~\eqref{eq:appOmegaReplicas} can be written as
\begin{equation}
   \E_\text{data}[\Omega^n] = \int \diff b \diff r\! \diff q_0\! \diff q \,
    \exp\! \left[\frac{n \beta L}{2} \left( G_S + \alpha^+ G_+ + \alpha^- G_- \right)\right]\,,
   \label{eq:theoryAction}
\end{equation}
where the functions of the order parameters $G_S$ (entropic term) and $G_\pm$ (energetic term) are detailed in the following. The remaining integral over the order parameters can be estimated for $L$ large through the saddle-point method.

\subsection{Entropic term}

First we evaluate the normalization of the measure $\mu(\vb{J})$ in Eq.~\eqref{eq:theoryJmeasure}. With our zero-sum conditions, compatible with the ones of~\citet{nadal1991potts}, this factor is given by
\begin{equation}
\begin{aligned}
    \Omega_0 
    &=  \int \prod_{i ,t} d J_i(t) \, \delta \Bigl[ \sum_{i, t} J^2_i(t) - L Q \Bigr] \prod_i \delta \Bigl[ \sum_{t} J_i(t)\Bigr]\\
    &=  \int \frac{d \hat{k}}{2\pi} \prod_{i} \frac{d g_i}{2 \pi} \prod_t d J_i(t) \, \exp \Bigl[ i \hat{k} \sum_{i, t} J^2_i(t) - L Q i \hat{k}+  i \sum_i g_i \sum_{t} J_i(t)\Bigr]\\
   &\sim \exp \Bigl[ \frac{L(Q-1)}{2} [1+\log (2 \pi )] -  \frac{L(Q-1)}{2} \log \Bigl(\frac{Q-1}{Q }\Bigr)-\frac{L}{2}\log (Q) \Bigr]\,,
\end{aligned}
\label{eq:PottsNormIntegral}
\end{equation}
as the integrals first over $\vb{J}$ and then over $\vb{g}$ are Gaussian and the large-$L$ result for the integral over $\hat{k}$ is obtained at the saddle point, which is located in $\hat{k} = i (Q-1)/(2Q)$.

In the RS case, the integral over the weights becomes
\begin{equation}
        I = \int \prod_{a,i} \frac{\diff g^a_i}{2\pi } \prod_{t} \diff J^a_i(t) \exp \Bigl[ -\frac{1}{2} \sum_{a,b} \sum_{t,u}  J^a_i (t) \Sigma^{ab}_{ij}(t,u) J^b_i (u)  - i\hat{r}\sum_{a,t} J^a_i(t) \delta_i(t)  + i  \sum_a g^a_i \sum_{t} J^a_i(t) \Bigr]\,,
    \label{eq:PottsRSJ}
\end{equation}
where the quadratic form $\Sigma$ is given by
\begin{equation}
     \Sigma^{ab}_{ij}(t,u) = \hat{k}  \delta_{ij} \delta_{tu} \delta_{ab}+ ( 2 \hat{q}_0 - \hat{q}) \delta_{ab} C_{ij}(t,u)  +  \hat{q} C_{ij}(t,u)
\end{equation}
(to write this equation, we rescaled all the hat variables by $-i$). The evaluation of this Gaussian integral is straightforward but tedious. It can be performed, for example, factorizing the double sums over replica indices via Gaussian linearization (Hubbard-Stratonovich transformation). The result for small $n$ is given by
\begin{equation}
    I = \exp \left\{ \frac{n}{2}\left[L(Q-1)\log(2\pi) - L \log\left( \frac{Q}{\hat{k}} \right) - \log \det A  
         - \hat{q} \tr(A^{-1} C) + \hat{r}^2 \bm{\delta}^\top A^{-1} \bm{\delta} \right]  \right\}\,,
    \label{eq:PottsRSA}
\end{equation}
where all the algebraic operations (determinants, traces, vector/matrix multiplications) are defined on the space spanned by Potts and input indices, the matrix $A$ is defined as
\begin{equation}
    A = \hat{k} \mathbb{I}_Q \otimes \mathbb{I}_L + \hat{w} C
\end{equation}
and $\hat{w} = 2 \hat{q}_0 - \hat{q}$. Notice that, even if the matrix $A$ is not invertible for $\hat{k}=0$ (the matrix $C$ has a spectrum with $L$ zero eigenvalues, corresponding to the eigenvectors $v^{(\ell)}$ with components $v^{(\ell)}_i(t) = \delta_{i\ell}1$), the combinations appearing in this formula are always well defined; in particular, the factor $L \log \hat{k}$ comes from the zero-sum conditions, which regularize the divergence in $\log \det A$.

Summing over the input dimensions and normalizing by~\eqref{eq:PottsNormIntegral} we get, for the function $G_S$ appearing in Eq.~\eqref{eq:theoryAction},
\begin{equation}
\begin{aligned}
    \frac{n \beta L G_S}{2} = \log \int_{-i\infty}^{i\infty} \frac{\diff \hat{k}}{4\pi} \frac{\diff \hat{w}}{2\pi} \frac{\diff \hat{q}}{2\pi} \frac{\diff \hat{r}}{2\pi}\, & e^{ \frac{n L }{2} \left[Q \hat{k}  + \hat{w} w +\hat{w} q   +\hat{q} w + 2\hat{r} r \right]  -\frac{n L (Q-1)}{2}\left[1 - \log \left(\frac{Q-1}{Q }\right) \right]}\\
          \times{} &  e^{-\frac{n}{2}\left[\log \det A - L \log \hat{k}
         + \hat{q} \tr(A^{-1}C) - \hat{r}^2 \bm{\delta}^\top A^{-1} \bm{\delta} \right]}\,,
\end{aligned}
\label{eq:appGS}
\end{equation}
 where $w = q_0 -q$.

\subsection{Energetic term}

\paragraph{SAT phase.}
When the problem is linearly separable (low density $\alpha^+$ and $\alpha^-$ of the classes, large distance $\bm{\delta}$ between their center), the perceptron at equilibrium reaches one of the many configurations corresponding to zero loss. This means that the optimization problem~\eqref{eq:minimization-Jb} becomes equivalent to a constraint statisfaction problem (CSP) where all the training data are required to be correctly classified by the decision boundary, namely in its satisfiable (SAT) phase. Formally, this means that the terms depending on the loss in Eq.~\eqref{eq:appOmegaReplicas} can be written as follows:
\begin{equation}
      e^{-\beta \sum_\mu (\kappa- y \Delta_{\mu}^a) \theta(\kappa -y \Delta_{\mu}^a )} \underset{\beta\to\infty}{\longrightarrow} \prod_\mu \theta(y\Delta_{\mu}^a - \kappa)\,.
\end{equation}
Thus, the part of the integral depending on the constraints from the training data reduces to the evaluation (respectively, $P$ and $N$ times) of the following two integrals
\begin{equation}
    I_\pm = \int \prod_{a}    \dfrac{\diff \Delta^a \diff \hat{\Delta}^a}{2 \pi} \theta(\pm \Delta^a - \kappa) e^{ -\frac{w}{2} \sum_{a} (\hat{\Delta}^a)^2 -\frac{q}{2} \left(\sum_{a} \hat{\Delta}^a\right)^2+i \sum_{a}  \hat{\Delta}^a \left( \Delta^a +b  \mp \frac{r}{2} \right) }\,.
\end{equation}
By Gaussian linearization,
\begin{equation}
    I_\pm = \int \Diff_q \xi \left[ \int   \dfrac{\diff \Delta \diff \hat{\Delta}}{2 \pi} \theta(\pm \Delta - \kappa) e^{ -\frac{w}{2}  \hat{\Delta}^2 +i   \hat{\Delta}  \left( \Delta +b + \xi \mp \frac{r}{2} \right) } \right]^n\,,
\end{equation}
where we defined the Gaussian measure as $\Diff_q \xi = \diff \xi \,\mathcal{N}(\xi|0,q)$. For small $n$, the final result is
\begin{equation}
    \log I_\pm \approx  \frac{n \beta}{2} G_\pm\,,
\end{equation}
with
\begin{equation}
\begin{aligned}
    \frac{\beta G_\pm}{2} & =  \int \Diff_q \xi \log\!\left[ H \!\left( \frac{\kappa \pm b \mp \xi - r/2}{\sqrt{w}} \right) \right]\,,\\
    H(x) & =  \frac{1}{2} \erfc\!\left(\frac{x}{\sqrt{2}}\right) = 1-H(-x)\,.
\end{aligned}
\label{eq:appGpmSAT}
\end{equation}
This equation defines the functions $G_\pm$ in Eq.~\eqref{eq:theoryAction} in the linearly separable phase.

\paragraph{UNSAT phase.} When the training set is not linearly separable, the corresponding CSP is in its unsatisfiable phase. In this case, the loss gives a non-trivial energetic contribution for each mis-classified input point. Calculations can be done as in the previous paragraph, by noticing that
\begin{equation}
    e^{-\beta (\kappa \mp \lambda ) \theta(\kappa \mp \lambda) } = e^{-\beta ( \kappa \mp \lambda ) } + \left[ 1- e^{-\beta (\kappa \mp \lambda )} \right] \theta(- \kappa \pm \lambda) \,.
\end{equation}
The result for the free energy contributions is
\begin{equation}
        \frac{\beta G_\pm}{2} = \int \Diff_q \xi \log \!\left\{\exp\!\left[ -\beta (\kappa   \pm b \mp \xi - r/2 ) + \frac{1}{2}\beta^2 w \right] H \!\left(- \frac{\kappa \pm b  \mp \xi - r/2 -\beta w}{\sqrt{w}} \right) + H \!\left(\frac{\kappa \pm b \mp \xi - r/2 }{\sqrt{w}} \right) \right\}.
        \label{eq:appGunsat1}
\end{equation}

We will assume in the following that the space of equilibrium configurations of the model shrinks to a single point as $\beta \to \infty$. As the variance of this space is given, in the replica symmetric framework, by the order parameter $w = q_0 - q$ (see Sec.~\ref{app:metrics}), for $\beta\to\infty$ we take the following scaling Ansatz:
\begin{equation}
\begin{aligned}
    w &\to \frac{x}{\beta}\,, &\quad q &\to q  \,, & \quad r &\to r \,,\\
    \hat{w} &\to \beta \hat{x} \,, & \hat{q} &\to \beta^2 \hat{q}\,, & \hat{r} &\to \beta \hat{r}\,,\quad & \hat{k} \to \beta \hat{k}\,,
\end{aligned}
\label{eq:appBetaScalings}
\end{equation}
where we re-defined all the order parameters to their infinite-$\beta$ value except $w$ and $\hat{w}$, for which we introduced the quantities $x$, $\hat{x}$ to avoid ambiguities. The scaling of all the others variables is fixed once 
$w$ scales as $1/\beta$ and we search for finite values in the limit.

We can push forward the analysis using the asymptotic expansion of the complementary error function, which gives for $\beta$ large
\begin{equation}
    H(\sqrt{\beta} a) \sim \theta(-a)+\frac{e^{-\frac{\beta a^2}{2}}}{\sqrt{2\pi \beta} a} \theta(a)\,,
\end{equation}
so that the logarithm in Eq.~\eqref{eq:appGunsat1} can be expressed as
\begin{equation}
    \log\{\cdots\!\} \sim \begin{cases}
        -\beta \left(\kappa   \pm b \mp \xi - r/2  - x/2 \right) & \text{if} \quad   \mp \xi > x + r/2 -\kappa \mp b \,,\\
        -\beta (\kappa   \pm b - r/2 \mp \xi)^2/(2x) & \text{if} \quad   \mp \xi < x+r/2 - \kappa \mp b \quad \land \quad    \mp \xi > r/2-\kappa \mp b \,,\\
        0 & \text{if} \quad  \mp \xi < r/2 - \kappa \mp b\,.
    \end{cases}
\end{equation}
As a result, we obtain Eq.~\eqref{eq:GpmUNSAT} for the functions $G_\pm$.

\subsection{Saddle-point equations for large \texorpdfstring{$L$}{L}}

For large $L$, the integral in Eq.~\eqref{eq:theoryAction} can be evaluated via the saddle-point method. The stationary points of the functions $G_S$, $G_\pm$ with respect to the parameters can be found by solving the following equations. We distinguish the two cases SAT/UNSAT we reported above.

\paragraph{SAT phase.}
We obtain stationary equations for the exponent in Eq.~\eqref{eq:theoryAction} by deriving with respect to all the order parameters and their conjugate variables the function at exponent in Eq.~\eqref{eq:appGS} and the functions in Eq.~\eqref{eq:appGpmSAT}.
Deriving with respect to the hat variables,
\begin{equation}
\begin{aligned}
    Q &= \hat{k} \left[ \frac{1}{L}  \tr( A^{-2})-\frac{1}{\hat{k}^2}  \right]   + \frac{\hat{w}}{L}  \tr(  A^{-2} C)  - \frac{\hat{q}}{L} \tr( A^{-2} C) +  \frac{\hat{r}^2}{L} \bm{\delta}^\top A^{-2} \bm{\delta} \,,\\
    q &= -  \frac{\hat{q}}{L} \tr(A^{-2} C^2) +  \frac{\hat{r}^2}{L} \bm{\delta}^\top A^{-2} C \bm{\delta}   \,,\\
    w  &=  \frac{\hat{k}}{L} \tr(A^{-2}C) +  \frac{\hat{w}}{L} \tr(A^{-2}C^2)   \,,\\
    r & = -  \frac{\hat{r}\hat{k}}{L} \bm{\delta}^\top A^{-2} \bm{\delta} -  \frac{\hat{r} \hat{w}}{L} \bm{\delta}^\top A^{-2}C \bm{\delta} \,,
\end{aligned}
\end{equation}
where we used the useful formulas
\begin{equation}
    A^{-1} = \hat{k} A^{-2} + \hat{w} A^{-2} C \,,\qquad
    A^2 = \hat{k}^2 \mathbb{I}_Q\otimes \mathbb{I}_L + 2 \hat{k} \hat{w} C + \hat{w}^2 C^2 \,.
\end{equation}
The equations obtained deriving with respect to the other set of parameters are
\begin{equation}
\begin{aligned}
    \hat{w} &= - \alpha_+ \partial_q (\beta G_+) - \alpha_- \partial_q  (\beta G_-) \,,\\
    \hat{w} + \hat{q} &= - \alpha_+ \partial_w (\beta G_+) - \alpha_-  \partial_w ( \beta G_-) \,,\\
    \hat{r} &= [- \alpha_+ \partial_r (\beta G_+) - \alpha_-  \partial_r (\beta G_-) ] /2 \,,\\
    0 &= \alpha_+ \partial_{b} (\beta G_+) + \alpha_-  \partial_{b}(\beta G_-) \,,
\end{aligned}
\end{equation}
with $\beta G_\pm$ given by~\eqref{eq:appGpmSAT}.

\paragraph{UNSAT phase.}
Using the scalings~\eqref{eq:appBetaScalings}, we can redefine the matrix $A \to A /\beta$, so that now
\begin{equation}
A = \hat{k} \mathbb{I}_Q\otimes \mathbb{I}_L + \hat{x} C\,.  
\end{equation}
At leading order in $\beta$, the stationary equations for the hat variables in the scaling regime become
\begin{equation}
    \begin{aligned}
    Q &=   -   \frac{\hat{q}}{L} \tr(A^{-2} C) +  \frac{\hat{r}^2}{L} \bm{\delta}^\top A^{-2} \bm{\delta} \,,\\
    q &= - \frac{\hat{q}}{L} \tr(A^{-2}C^2) +  \frac{\hat{r}^2}{L} \bm{\delta}^\top A^{-2} C \bm{\delta} \,,\\
    x &=   \frac{\hat{k}}{L} \tr(A^{-2} C) + \frac{\hat{x}}{L} \tr(A^{-2} C^2)  \,, \\
    r &= -  \frac{\hat{r}\hat{k}}{L} \bm{\delta}^\top A^{-2} \bm{\delta} -  \frac{\hat{r}\hat{x}}{L} \bm{\delta}^\top A^{-2} C \bm{\delta} \,,
\end{aligned}
\end{equation}
while the other set of equations is
\begin{equation}
    \begin{aligned}
    \hat{x} &= - \alpha_+ \partial_q G_+ - \alpha_- \partial_q  G_- \,,\\
    \hat{q} &= - \alpha_+ \partial_x G_+ - \alpha_-  \partial_x  G_- \,,\\
    \hat{r} &= [- \alpha_+ \partial_r G_+ - \alpha_-  \partial_r G_-) ] /2 \,,\\
    0 &= \alpha_+ \partial_{b}  G_+ + \alpha_-  \partial_{b}G_- \,,
\end{aligned}
\end{equation}
with $G_\pm$ given by Eq.~\eqref{eq:GpmUNSAT}. These last equations admit a semi-analytical expression for the derivatives:
\begin{equation}
\begin{aligned}
    \partial_x G_\pm ={}& - \frac{G_\pm}{x} + \frac{x - \mathcal{K}_\pm }{x}\erfc\left(\frac{x - \mathcal{K}_\pm}{\sqrt{2q}}\right) - \frac{2}{x}\sqrt{\frac{q}{2\pi}} e^{-\frac{(\mathcal{K}_\pm-x)^2}{2 q}}\,,\\
    \partial_{q} G_\pm ={}& -\frac{1}{2 x}  \left[\erfc\left(-\frac{\mathcal{K}_\pm}{\sqrt{2 q} }\right)-\erfc\left(\frac{x - \mathcal{K}_\pm}{\sqrt{2q}}\right)\right]\,,\\
    \partial_{r} G_\pm ={} & \frac{\mathcal{K}_\pm}{2 x}   \erfc\left(-\frac{\mathcal{K}_\pm}{\sqrt{2 q}}\right)+\frac{x - \mathcal{K}_\pm}{2x} \erfc\left(\frac{x - \mathcal{K}_\pm}{\sqrt{2q}}\right) 
    +\frac{1}{x}  \sqrt{\frac{q}{2\pi}}  \left(e^{-\frac{\mathcal{K}_\pm^2}{2 q}} - e^{-\frac{(\mathcal{K}_\pm-x)^2}{2 q}}\right)\,,\\
    \partial_{b} G_\pm  ={}& \mp \frac{\mathcal{K}_\pm}{ x}  \erfc\left(-\frac{\mathcal{K}_\pm}{\sqrt{2 q}}\right) \mp \frac{x - \mathcal{K}_\pm}{x}  \erfc\left(\frac{x - \mathcal{K}_\pm}{\sqrt{2q}}\right)
    \mp \frac{2}{x}  \sqrt{\frac{q}{2\pi}}  \left(e^{-\frac{\mathcal{K}_\pm^2}{2 q}} - e^{-\frac{(\mathcal{K}_\pm-x)^2}{2 q}}\right)\,.
\end{aligned}
\label{eq:gradGpmUNSAT}
\end{equation}

\paragraph{SAT/UNSAT surface.}
The equations above can be used to draw the phase diagram of the model with respect to the external control parameters $\alpha^\pm$, $\kappa$ (and, possibly, to the vectors $\vb{M}$, $\bm{\delta}$). In this space, the SAT/UNSAT transition surface expressing the critical value of one of these control parameters as a function of the others can be plotted requiring that $w \to 0$ (approaching the transition from the SAT phase) or $x\to \infty$ (approaching the transition from the UNSAT phase).

\subsection{Ising case (\texorpdfstring{$Q=2$}{Q=2}) \label{sec:app_ising_theory}}

The case $Q=2$ (Ising configurations, with the index $t=\pm 1$) admits a much simpler solution that dates back to~\citet{gardner1988space},~\citet{monasson1992}. In this case, the probability distribution of the data can be parametrized as
\begin{equation}
\begin{aligned}
     &M_i(\pm 1) = \frac{1\pm m_i}{2}\,,\qquad \delta_i (\pm 1) = \pm \frac{\delta_i}{2}\,, \\
     &C_{ij}(+1,+1) = C_{ij}(-1,-1) = - C_{ij}(+1,-1) = -C_{ij}(-1,+1) = \frac{\Gamma_{ij}}{4}\,,
\end{aligned}
    \end{equation}
where now $m_i$, $\delta_i$, $\Gamma_{ij}$ are scalars. The saddle-point equations for the hat variables become
\begin{equation}
    \begin{aligned}
    1 &=   -   \frac{\hat{q}}{L} \tr(A^{-2} \Gamma) +  \frac{\hat{r}^2}{L} \bm{\delta}^\top A^{-2} \bm{\delta} \,,\\
    q &= - \frac{\hat{q}}{L} \tr(A^{-2} \Gamma^2) +  \frac{\hat{r}^2}{L} \bm{\delta}^\top A^{-2} \Gamma \bm{\delta} \,,\\
    x &=   \frac{\hat{k}}{L} \tr(A^{-2} \Gamma) + \frac{\hat{x}}{L} \tr(A^{-2} \Gamma^2)  \,, \\
    r &= -  \frac{\hat{r}\hat{k}}{L} \bm{\delta}^\top A^{-2} \bm{\delta} -  \frac{\hat{r}\hat{x}}{L} \bm{\delta}^\top A^{-2} \Gamma \bm{\delta} \,,
\end{aligned}
\label{eq:app_sp_ising_unsat}
\end{equation}
where now
\begin{equation}
    A = \hat{k} \mathbb{I}_L + \hat{x} \Gamma\,,
\end{equation}
The other set of equations remains unchanged and the case of the SAT phase can be obtained as before.

\subsection{SVM\label{app:SVM}}

The case of the soft-margin SVM can be cast in its standard form (see Sec.~\ref{app:classifiers}) by starting from the loss defined in Eq.~\eqref{eq:loss}, and scaling $\vb J \to \kappa \vb J$, $b \to \kappa b$, $\lambda \to \tilde{\lambda}/\kappa$, obtaining the optimization problem (we now move the L2 regularization from the measure~\eqref{eq:theoryJmeasure} to the training energy, as this is the standard formulation for SVMs)
\begin{equation}
    (\vb{J}^\star,b^\star) = \underset{\vb J \in \mathcal{S},b \in \mathbb{R}}{\arg \min}  \left\{ \sum_{\mu = 1}^P \max \left[ 0 , 1 - \Delta^+_\mu (\vb J, b)\right] + \sum_{\nu = 1}^N \max \left[0 , 1 + \Delta^-_\nu (\vb J, b)\right]
    + \tilde{\lambda} \frac{ \lVert \vb{J} \rVert^2_2}{2} \right\}\,.
    \label{eq:minimization-SVM}
\end{equation}
From this, we can define the associated finite-temperature statistical mechanics model from the SVM measure in Eq.~\eqref{eq:theoryJmeasure} and proceed as before. The calculation is the same, provided that we take $\kappa \to 1$, we substitute $\hat{k}$ from~\eqref{eq:PottsRSJ} with $\beta\tilde{\lambda}$, and we do not integrate over it (being now fixed as an external hyperparameter). The scaling of $\tilde{\lambda}$ with $\beta$ mirrors the fact that, at finite regularization, the SVM is finding a single solution of the optimization problem~\eqref{eq:minimization-SVM} even in the linearly separable case, the \emph{max-margin solution}, so that the scalings~\eqref{eq:appBetaScalings} should be enforced in the whole phase diagram.

\section{Details on the theoretical derivation of training and generalization metrics\label{app:metrics}}

\begin{table}[t]
\caption{Additional common performance metrics that can be computed from Table~\ref{tab:metrics}}
\label{tab:app_metrics}
\vskip 0.15in
\begin{center}
\begin{small}
    \begin{tabular}{lcr}
    \toprule
        $\text{NPV}$  & $ \varphi^- \text{TNR}(0)/(\varphi^- \text{TNR}(0) + \varphi^+ \text{FNR}(0))$ & Negative predicted value\\ \addlinespace
        $\text{FDR}$  & $ 1-\text{PPV}$ & False discovery rate\\ \addlinespace
        $\text{FOR}$  & $ 1-\text{NPV}$ & False omission rate\\ \addlinespace
        $\text{MCC}$  & 
        $ \sqrt{\text{TPR}(0)\cdot \text{TNR}(0) \cdot \text{PPV} \cdot  \text{NPV}} - \sqrt{\text{FNR}(0)\cdot\text{FPR}(0) \cdot \text{FOR} \cdot \text{FDR}}$
        & Matthews correlation coefficient\\ \addlinespace
        $\text{F}_1$  & $ 2 \cdot \text{PPV}\cdot\text{TPR}(0)/(\text{PPV} + \text{TPR}(0))$ & F$_1$ score\\ \addlinespace
        $\text{FM}$  & $ 2 \cdot\sqrt{\text{PPV}\cdot\text{TPR}(0)}$ & Fowlkes–Mallows index\\
        \bottomrule
    \end{tabular}
\end{small}
\end{center}
\vskip -0.1in
\end{table}

\begin{figure}
    \centering
    \includegraphics[width=\textwidth]{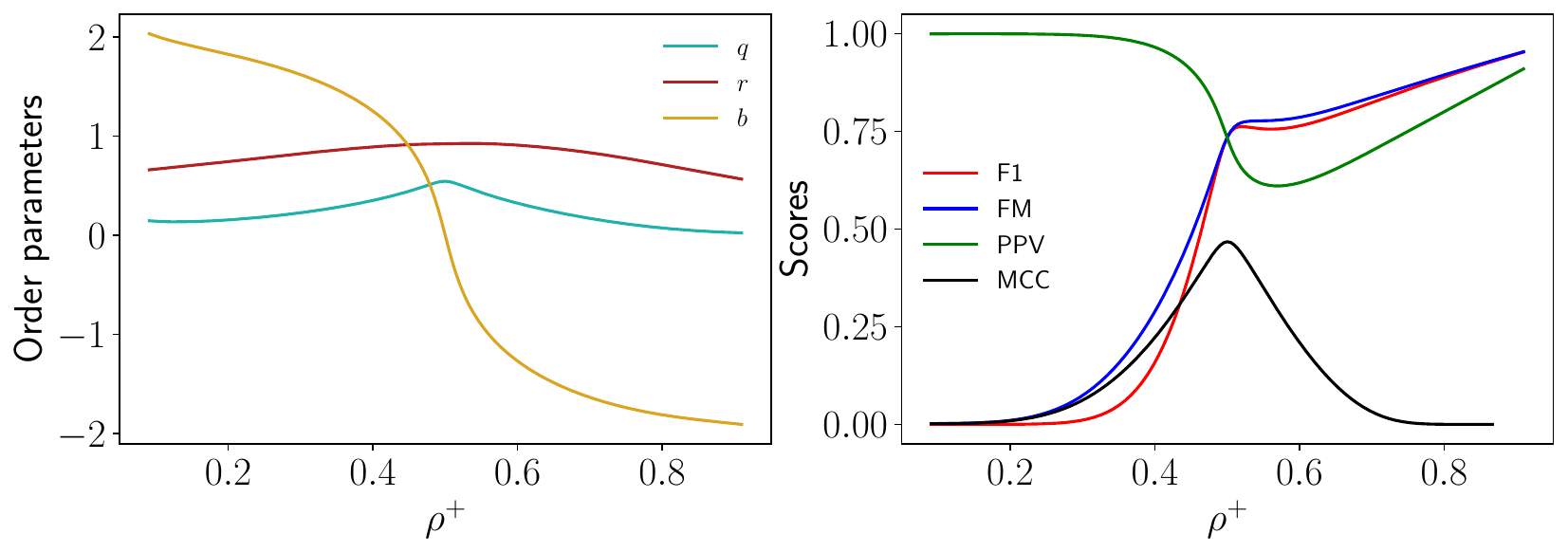}
    \put(-480,175){{\bf a}}
    \put(-243,175){{\bf b}}
    \caption{\label{fig:app_metrics_behavior} {\bf a)} Behaviors of the order parameters $q$, $r$, $b$ for the data statistics in \cref{fig:fig_theory}a. The bias $b$ becomes steep around $\rho^{+} \sim 0.5$ and this effect is more evident the larger the value of $\kappa$ is. {\bf b)} Additional metric behaviors defined in this work (see \cref{tab:app_metrics,,tab:metrics}), for synthetic data as in \cref{fig:fig_theory}a.} 
\end{figure}

The order parameters introduced in Sec.~\ref{sec:theo1} and obtained from the replica method explained above have a clear interpretation in terms of the low-order statistics -- with respect to thermal fluctuations and quenched disorder --  of the collective variables $\Delta^y$ ($y \in \{ -1, +1 \}$) after training. We recall here the definition of $\Delta^y$ for convenience
\begin{equation}
    \Delta^y (\vb J, b) = \sum_{i,t} \frac{J_i(t) s_{i}^{y}(t)}{\sqrt{L}}  - b\,,
\end{equation}
where $(\vb{J},b)$ is an equilibrium configuration of the canonical ensemble described by the partition function~\eqref{eq:Omega} and $\vb{s}^y$ is an input point from the probability distributions described by Eq.~\eqref{eq:Data}. We need to distinguish the two cases of $\vb{s}^y$ being part of the training set or $\vb{s}^y$ being part of the test set (\ie{}, independent from -- but identically distributed as -- the training set), denoting $\Delta^y_{\text{train}}$ and $\Delta^y_{\text{test}}$ the corresponding collective variables. It can be shown that their distribution for $L$ large is given by
\begin{align}
   p(\Delta^y_{\text{train}} |\xi)  &=  \frac{e^{-\beta \ell (y , \Delta_{\text{train}}^y)} \mathcal{N}\!\left(\Delta_{\text{train}}^y \!\left|  \xi - b +y r/2 ,  q_0 - q \right. \right)}{ {\displaystyle  \int } \!\diff \Delta \, e^{-\beta \ell (y , \Delta)} \mathcal{N}\!\left(\Delta \!\left|  \xi - b +yr/2 ,  q_0 - q \right. \right) },
\label{eq:DeltaDistrTrain}\\
p(\Delta^y_{\text{test}} |\xi)  &=  \mathcal{N}\!\left(\Delta_{\text{test}}^y \!\left|  \xi - b +yr/2 , q_0 - q \right. \right),
\label{eq:DeltaDistrTest}
\end{align}
where $\xi$ is an auxiliary normal random variable, in both cases and for both classes distributed as
\begin{equation}
p(\xi) = \mathcal{N}(\xi | 0, q)\,.    
\end{equation}
Note that in the test case the marginal of $\Delta$ can be obtained explicitly, as
\begin{equation}
    p(\Delta^y_{\text{test}} )  =  \mathcal{N}\!\left(\Delta_{\text{test}}^y \!\left| \frac{yr}{2} - b , q_0 \right. \right)\,,
\end{equation}
which reduces to Eq.~\eqref{eq:DeltaDistr} for large $\beta$ and in the UNSAT phase where $q_0 = q + x/\beta$. The Gaussian nature of $\Delta^y_{\text{test}}$ is due to the fact that the weights learned by the classifier and the points in the test set are independent, and from central limit theorems.
Eq.~\eqref{eq:DeltaDistrTrain} and~\eqref{eq:DeltaDistrTest} are classical results from the statistical mechanics of disordered systems, and can be obtained by mapping the replica approach we devised so far to the cavity method from~\citet{mezard1989cavity}. For a more recent derivation, see~\citet{agoritsas2018cavity}; for a way to obtain the train distribution directly from the replica approach, see~\citet{abbott1989domains}, extended in a more general non-convex setting in~\citet{urbani2017jamming}.

Through these probability distributions, we obtain theoretical predictions for training and generalization metrics; in particular, train error and mis-classification error (corresponding to $1 - \text{ACC}$) follow from
\begin{equation}
    \mathcal{E}_\text{train.}(\rho^+,\rho^-) =  \sum_{y \in \{\pm \} } \rho^y \braket{  \theta( - y \Delta^y_\text{train} ) }_{\Delta^y_\text{train}}\,, \qquad
    \mathcal{E}_{\text{gen.}}(\varphi^+, \varphi^-) = \sum_{y \in \{\pm \} } \varphi^y \braket{  \theta( - y \Delta^y_\text{test} ) }_{\Delta^y_\text{test}} \,,
    \label{eq:Errors}
\end{equation}
while training and generalization losses are given by
\begin{equation}
    \mathcal{L}_\text{train.}(\rho^+,\rho^-) =  \sum_{y \in \{\pm \} } \rho^y \braket{  \ell(y, \Delta^y_\text{train} ) }_{\Delta^y_\text{train}}\,, \qquad
    \mathcal{L}_{\text{gen.}}(\varphi^+, \varphi^-) = \sum_{y \in \{\pm \} } \varphi^y \braket{  \ell(y, \Delta^y_\text{test} ) }_{\Delta^y_\text{test}} \,.
    \label{eq:Losses}
\end{equation}
Alternatively, the training loss can be obtained in a more simple way, as the $\beta\to \infty$ limit of the free-energy per-sample of the physical model, so that from Eq.~\eqref{eq:logOmega} we directly obtain
\begin{equation}
    \mathcal{L}_\text{train.}(\rho^+,\rho^-) = \frac{1}{2(\alpha^+ + \alpha^-)} ( G_S+\alpha^+ G_+ + \alpha^- G_-)\,,
\end{equation}
where the functions of the order parameters are evaluated on their saddle-point values. Other common performance metrics that can be computed in this way can be found in Tables~\ref{tab:metrics} and~\ref{tab:app_metrics}, from Proposition~\ref{propERM}.

\subsection{Convergence rate of the mis-classification error \texorpdfstring{$\mathcal{E}_{\text{gen.}}$}{Egen}\label{sec:EgenConv}}

Here we fully derive the analytical expression for the balanced mis-classification error as a function of the composition of the training set, in the limit where the training set size is large. To obtain the expression in Eq.~\eqref{eq:misclassification_error_largeK}, we work in the Ising case ($Q=2$) using the notation of section Sec.~\ref{sec:app_ising_theory}, setting $m_i = m$, $\Gamma_{ij} = \delta_{ij}(1 -m^2)$. In this scenario the saddle point equations are simpler and can be carried out without any numerical support. In particular, for the hat variables the system \eqref{eq:app_sp_ising_unsat} becomes
\begin{equation}
\label{eq:app_sp_ising_unsat2}
    \begin{aligned}
        1  &= \dfrac{\hat{r}^2 u^2 -\hat{q} (1-m^2)}{[\hat{k} + \hat{x} (1-m^2)]^2} \,, \quad&
        x  &= \dfrac{1-m^2}{\hat{k} + \hat{x} (1-m^2)}\,, \\
        \frac{q}{1-m^2} &=  \dfrac{\hat{r}^2 u^2 - \hat{q} (1-m^2)}{[\hat{k} + \hat{x} (1-m^2) ]^2} ,\quad &
        r &= - \dfrac{\hat{r} u^2 }{\hat{k} + \hat{x} (1-m^2)},
    \end{aligned}
\end{equation}
where we defined the quantity $u = \lVert \bm{\delta} \rVert_2/\sqrt{L}$. The system is overdetermined due to the choice of a position-independent vector $m_i = m $. Indeed, the constraint enforced by the definition of $q_0$ is equivalent to the spherical constraint on the weights $J_i$, hence here $q_0^\star = (1-m^2)$; consequently, the conjugate variable $\hat{q}_0$ plays no role and we can set $\hat{q}_0 = 0$, from which $\hat{x} = - \hat{q}$  follows. The system of equation~\eqref{eq:app_sp_ising_unsat2} gives
\begin{equation}
    \begin{aligned}
        q^\star &= 1-m^2\,, \quad & 
        \hat{k} &= \dfrac{q^\star}{x} +\hat{q} q^\star\,,\\
        \hat{q} &= \dfrac{q^\star}{x^2} \left( \dfrac{r^2}{u^2} - 1 \right)\,, \quad &
        \hat{r} &= - \dfrac{r q^\star}{x u^2 }\,.
    \end{aligned}
\end{equation}
Replacing the hat variables in the entropic contribution, we have to perform the extremization over $(x,r,b)$ of the training loss
\begin{equation}
\label{eq:app_qtrain_ising_unsat}
    \mathcal{L}_{\text{train.}} (\alpha^+, \alpha^-) = \dfrac{1}{2 (\alpha^+ + \alpha^-)}\extr_{x, r, b}  \left[\dfrac{q^\star}{x u^2 } \left( u^2 - r^2 \right) + \alpha^+ G_+ + \alpha^- G_- \right],
\end{equation}
where $G_{\pm}$ are given by Eq.~\eqref{eq:GpmUNSAT} with $q = q^\star $. In addition, we fix $\alpha^+$ and minimize the training loss over the number of negative samples $\alpha^-$, to obtain the optimal imbalance ratio in terms of training performances. Now we first derive Eq.~\eqref{eq:app_qtrain_ising_unsat} w.r.t. to $x$ obtaining 
\begin{equation}
\label{eq:app_eqforx}
\begin{aligned}
    -\dfrac{1}{x^2} \Bigg\{
    \dfrac{q^\star (u^2-r^2)}{u^2}  &+ \sum_{y = \pm} \alpha^y \sqrt{\dfrac{q}{2 \pi}}\left[ (\mathcal{K}_y -x) e^{-\frac{1   }{2q} (x - \mathcal{K}_y)^2} - \mathcal{K}_y e^{-\frac{1}{2q} \mathcal{K}_y^2}\right] \\
    &- \sum_{y=\pm} \alpha^y \left[ (\mathcal{K}_y^2 - x^2 + q ) \erfc \left( \dfrac{x - \mathcal{K}_y}{\sqrt{2q}} \right) - (\mathcal{K}_y^2 + q) \erfc \left( - \dfrac{\mathcal{K}_y}{\sqrt{2q}} \right) \right] \Bigg\} = 0.
\end{aligned}
\end{equation}
We discard the solution $x \to \infty$ as it is only valid when approaching the SAT/UNSAT transition, and we call $x = x^\star (r,b,\alpha^+,\alpha^-)$ the other solution. The stationary conditions of $\mathcal{L}_{\text{train.}}$ in Eq.~\eqref{eq:app_qtrain_ising_unsat} using Eq.~\eqref{eq:app_eqforx} w.r.t. $(\alpha^-, r,b)$ are
\begin{equation}
    \begin{aligned}
    \alpha^- :&  & \dfrac{\alpha^+}{2(\alpha^+ + \alpha^-)^2} \left[  
    (x-\mathcal{K}_-) \erfc \left( \dfrac{x - \mathcal{K}_-}{\sqrt{2q}} \right) - (x-\mathcal{K}_+) \erfc \left( \dfrac{x - \mathcal{K}_+}{\sqrt{2q}} \right) \right] &= 0, \\ \\
    r :&  & \left[ \pdv{x^\star(r,b)}{r} - \dfrac{1}{2} \right]  \sum_{y = \pm} \alpha^y \left[ \erfc \left( \dfrac{x - \mathcal{K}_y}{\sqrt{2q}} \right) -\dfrac{2 (x - \mathcal{K}_y)}{\sqrt{2q \pi}} \exp \left\{-\dfrac{(x-\mathcal{K}_y)^2}{2q}\right\} \right]  &= 0, \\ \\
    b :&  &  \sum_{y = \pm} \alpha^y \left( \pdv{x^\star(r,b)}{r} - y \right) \left[ \erfc \left( \dfrac{x - \mathcal{K}_y}{\sqrt{2q}} \right) -\dfrac{2 (x - \mathcal{K}_y)}{\sqrt{2q \pi}} \exp \left\{-\dfrac{(x-\mathcal{K}_y)^2}{2q} \right\} \right]  &=0 .
    \end{aligned}
\end{equation}
The first equation gives the condition $\mathcal{K}_- = \mathcal{K}_+$, yielding $b^\star=0$. In the large limit $\kappa \gg 1$, where we have shown to achieve the absolute best generalization performance, we get $\mathcal{K}_{\pm} \to \infty$. Hence Eq.~\eqref{eq:app_eqforx} admits the explicit solution
\begin{equation}
   x^\star =\dfrac{1}{u} \sqrt{\dfrac{q^\star (u^2-(r^\star)^2)} {\alpha^+ + \alpha^-} } \,,
\end{equation}
with 
\begin{equation}
    r^\star = u\sqrt{\dfrac{(\alpha^+ + \alpha^-) u^2} {4q^\star + u^2 (\alpha^+ + \alpha^-)} }\,, \qquad
    \alpha^- = \alpha^+\,.
\end{equation}
We conclude that optimal classification over the training set is reached at balanced number of samples, for high margin and infinite dataset size.
For large $\alpha^{\pm}$, we finally derive the convergence rate of the mis-classification error in the balanced case ($\varphi^\pm = 0.5$), as 
\begin{equation}
    \mathcal{E}_{\text{gen.}} \underset{\alpha^{\pm} \gg 1}{\sim} \dfrac{1}{2} \erfc{\! \left( \dfrac{\normLtwonosq{\bm{\delta}}}{2 \sqrt{2L(1-m^2)}}\right)} \delta( \alpha^{+} - \alpha^{-} ) + \dfrac{\mathcal{C}}{\alpha^+ + \alpha^-} \delta( \alpha^{+} - \alpha^{-} ) +O ((\alpha^{\pm}) ^{-2} ),
\end{equation}
where 
\begin{equation}
  \mathcal{C} = \sqrt{1-m^2} \exp \left\{ - \dfrac{\normLtwo{\bm{\delta}}}{8L(1-m^2)}\right\}.
\end{equation}

\section{Details on the theory of oversampling strategy}
Here we provide details about the main steps to adapt the analytical computation of Sec.~\ref{app:replicas} when the minority class is oversampled $c$ times. We assume, as in most of the paper, that the minority class is the positive one. The mathematical formulation in this setting follows immediately from Eq.~\eqref{eq:OmegaReplicated} by introducing the oversampling factor $c$, accounting for the multiplicity of each positive sample:
\begin{equation}
      \exp \left\{-\beta \sum_{\mu=1}^{\alpha^+\! L} (\kappa-\Delta_{\mu}^a) \theta(\kappa -\Delta_{\mu}^a ) \right\} \quad \longrightarrow \quad \exp \left\{-\beta c  \sum_{\mu=1}^{\alpha^+\! L} (\kappa-\Delta_{\mu}^a) \theta(\kappa -\Delta_{\mu}^a ) \right\}.
\end{equation}
In the phase where the dataset is linearly separable, oversampling the positive class is pointless, as the perceptron already correctly classifies all the training data and adding equivalent ones does not yield any benefit. On the other hand, in the UNSAT phase the presence of the factor $c$ becomes relevant and affects the energetic contribution from the positive class, see Eq.~\eqref{eq:GpmUNSAT}; in practice, the quantities $\beta$, $x$ are multiplied by the degree of oversampling $c$. The positive function $G_+$ becomes 
\begin{equation}
        G_+
    = -\sqrt{\frac{q}{2\pi }} \left( \frac{cx - \mathcal{K}_+ }{cx} e^{-\frac{(cx - \mathcal{K}_+)^2}{2 q}}+ \frac{\mathcal{K}_+}{cx}  e^{-\frac{\mathcal{K}_+^2}{2 q}}
      \right) 
    +\frac{(cx - \mathcal{K}_+)^2+q}{cx} H\!\left(\frac{cx-\mathcal{K}_+}{\sqrt{q}}\right)
    -\frac{\mathcal{K}_+^2+q}{cx} H\!\left(\frac{-\mathcal{K}_+}{\sqrt{q}}\right)
\end{equation}
and all its derivatives in Eq.~\eqref{eq:gradGpmUNSAT} are changed accordingly. The procedure to solve the saddle-point equations remains analogous.

\section{Oversampling vs undersampling: the choice depends on the performance metric}
In \cref{sec:sampling_strategies} we discussed the advantage of restoring balance in the dataset, oversampling or undersampling, evaluating the generalization performances in terms of the BA metrics. However, same data distributions and sizes can lead to different optimal strategies depending on how performances are evaluated, \ie{} on the specific performance metric adopted. We discuss this phenomenon within our theoretical framework, comparing the mis-classification error $\mathcal{E}_{\text{gen.}}$ and a distance-based metric, namely the generalization loss $
\mathcal{L}_{\text{gen.}}$ defined in Eq.~\eqref{eq:Losses}. Both metrics are such that the lower the better, as they measure wrongly classified data points. 

\begin{figure*}[t]
    \includegraphics[width=\textwidth]{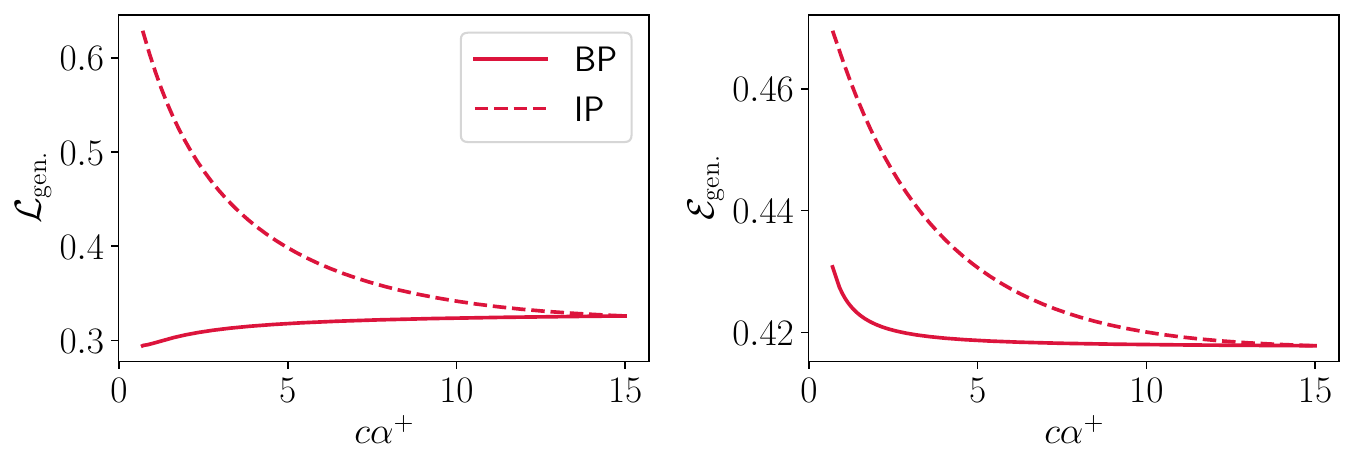}
    \caption{\label{fig:app_choice_metrics} Comparison between the mis-classification error and generalization loss as a function of the actual minority training size (including oversampling) $c \alpha^+$ for protocols IP and BP. Here $L=100$, $Q=2$, $\kappa = 0.4$, $\alpha^- = 15$, $\alpha^+ = 0.5$ with $m_i = 0.4$ over each position, $\delta_i$ are i.i.d. as $\mathcal{N}(0.625,0.75)$ and covariances are diagonal. Note that for these metrics the lower the better.}
\end{figure*}

The so-defined generalization loss gives the average distance of the mis-classified examples from the decision boundary set by the classifier. In practice, low $\mathcal{L}_{\text{gen.}}$ and high $\mathcal{E}_{\text{gen.}}$ values correspond to many mis-classified examples close to the decision boundary, while the opposite refers to situations where there are few mis-classified examples far away from the decision boundary. Using the same notation as in \cref{sec:sampling_strategies} with $c \in [1, \alpha^{-}/\alpha^{+}]$, we compare the metrics $\mathcal{E}_{\text{gen.}}$ and $\mathcal{L}_{\text{gen.}}$ under two different protocols, defined as follows
\begin{enumerate}
    \item[(IP)]  {\it \hypertarget{IP}{Imbalanced Protocol}}: we leave the majority class as it is and augment the minority class size, thus performing training with $c \alpha^+$ positives and $\alpha^-$ negatives;
    \item[(BP)] {\it \hypertarget{BP}{Balanced Protocol}}: we undersample the negatives down to the same size of positives, hence dealing with a balanced training set of total size $2c \alpha^+$. Note that this is the protocol adopted elsewhere in the work and it is the only one restoring balance between classes.
\end{enumerate}
We show results in \cref{fig:app_choice_metrics}: in the same conditions, the optimal strategy minimizing the generalization loss is the BP protocol with $c=1$ (\ie{}, random undersampling of the majority class), while the optimal one for the mis-classification error is the BP protocol with $c = \alpha^-/\alpha^+$ (\ie{}, random oversampling of the minority class).

\section{Lattice Proteins \label{sec:app_LP}}
In many real-world applications tokenization and categorical variables are routinely encountered in machine learning. For example, biological applications are one of such cases, where Potts indices run over the amino acid or nucleotides alphabet, \ie{} $Q=20$ or $Q=4$ respectively. In this section we report an analysis carried out on synthetic protein data -- namely Lattice Proteins -- where we aim to distinguish between two Lattice Proteins in a bounded or unbounded state. In this analysis we note that our theory quantitatively predicts numerical results even if we ignore the covariance matrix, \ie{} $C_{ij}(t,u) = [M_i^2(t) - M_i(t) M_i(u)] \delta_{ij}$; in other terms, the first-order statistics of the data is sufficient to reproduce numerical results, despite the data contain non-trivial correlations.

\label{app:lattice_prot}
Lattice Proteins (LPs) models consist of artificial protein sequences used to reproduce relevant features of real proteins and investigate their folding properties \citep{li1996emergence,mirny2001protein}. A LP is defined as a self-avoiding path of over a $3 \times 3 \times 3$ lattice cube. There are 103,406 possible configurations (structures) over the lattice cube, excluding global symmetries \citep{shakhnovich1990enumeration}. A sequence $\vb{v}$ is given by the chain of $L=27$ amino acids sitting on each site of the lattice cube. A structure $S$ is defined by its contact matrix $\vb{c}^{S}$ such that - given two sites $i$,$j$ on the lattice cube
\begin{equation}
\label{eq:cmap}
\begin{split} 
c_{ij}^S = \begin{cases}
1  \qquad i,j \quad \text{in contact}, \\
0 \qquad \text{otherwise}.
\end{cases}
\end{split}
\end{equation} 
The probability that sequence $\vb{v}$ folds into structure $S$ is
\begin{equation} 
\label{eq:app_pnat} 
p_{\text{nat}}(S |\vb{v} ) = \dfrac{e^{- E(\vb{v} | S)}}{\sum_{S'=1}^{R} e^{- E(\vb{v} | S')} },
\end{equation}
where $R$ is a representative subset of all possible structures ($R ={}$10000 in this case). The energy of the sequence in a structure $E(\vb{v} | S)$ is given by 
\begin{equation} 
\label{eq:app_En_MJ}
E(\vb{v} | S) = \sum_{i<j} c_{ij}^S E_{MJ}(v_i,v_j),
\end{equation}
where residues in contact interact via the Miyazawa-Jernigan matrix $E_{MJ}$, a proxy containing effective interaction energies for each amino acid pair. Eventually we let two sequences $\vb{v}_1$, $\vb{v}_2$ -- folded in structure $S_1$, $S_2$ respectively -- interact to form a unique compound via the interaction energy
\begin{equation} 
\label{eq:app_Int_MJ}
I(\vb{v}_1,\vb{v}_2 | S_1+S_2, \pi ) = \sum_{i<j} c_{ij}^{(S_1+S_2, \pi)} E_{MJ}(v_i,v_j) ,
\end{equation}
where now the contact matrix $\vb{c}^{(S_1+S_2)}$ is defined over both structures and the sum runs over all sites of both lattice cubes.  The index $\pi$ in Eq.~\eqref{eq:app_Int_MJ} labels a specific orientation of the interaction. In analogy with \cref{eq:app_pnat}, the probability that sequences $\vb{v}_1,\vb{v}_2$ folded into the compound $S_1 + S_2$ interact is 
\begin{equation} 
\label{eq:app_pint} 
p_{\text{int}}(\pi = 0, S_1 + S_2 |\vb{v}_1,\vb{v}_2 ) = \left[ \dfrac{e^{- I(\vb{v}_1,\vb{v}_2 | S_1+S_2,\pi=0 ) }}{\sum_{\pi'=1}^{R'} e^{- I(\vb{v}_1,\vb{v}_2 | S_1+S_2,\pi')} } \right]^{\chi},
\end{equation}
where $R'$ is the total number of orientations ($R' = {}$144 in this case) and we added the exponent $\chi$ to tune the interaction strength. Depending on the value of $\chi$, we refer to the compound $S_1 + S_2$ as bounded ($\chi=5$), mildly bounded ($\chi=0.1$) or unbounded ($\chi=0$).

\begin{figure*}[t]
    \centering
    \includegraphics[width=0.44\textwidth]{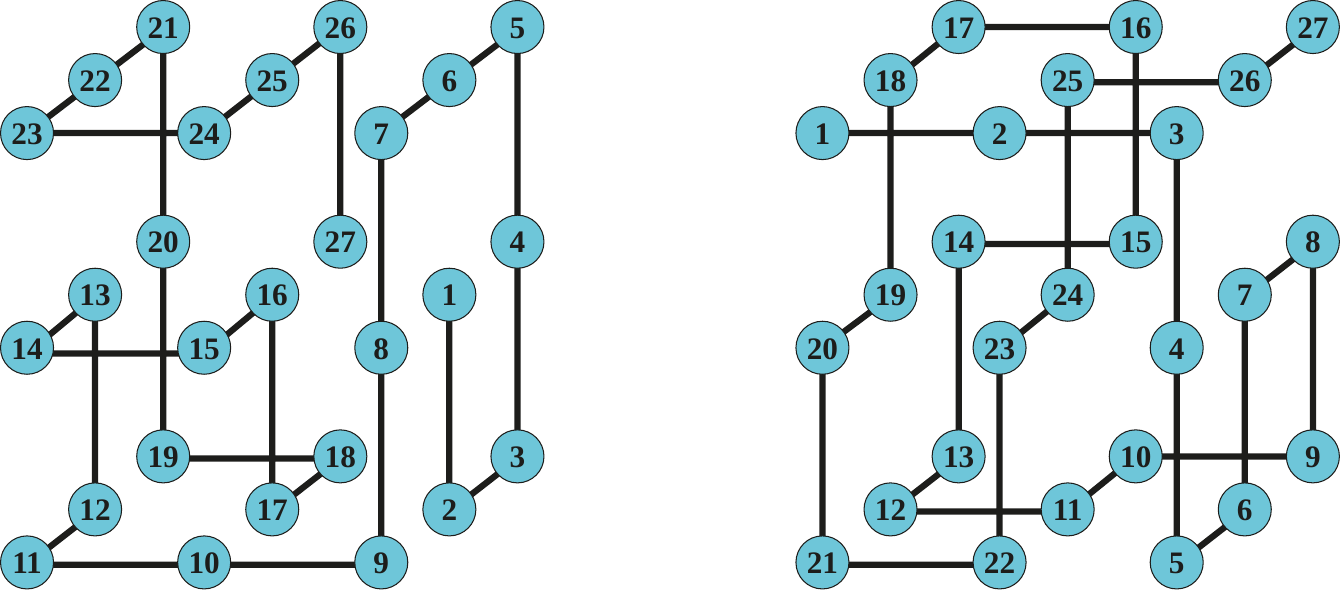}
        \hspace{.1cm}
    \includegraphics[width=0.52\textwidth]{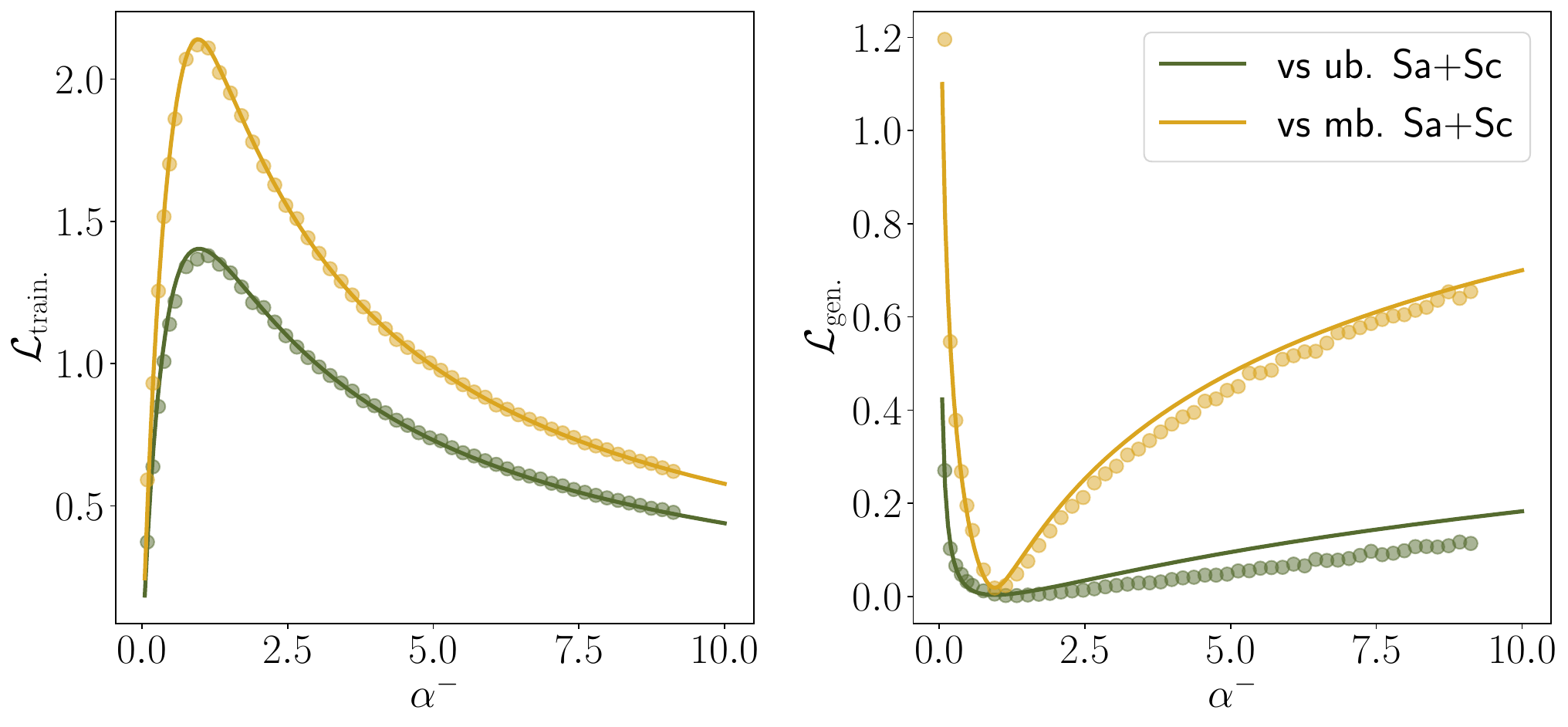}
    \put(-480,125){\bf a}
    \put(-255,125){\bf b}
    \put(-440,110){$S_C$}
    \put(-312,110){$S_A$}

    \caption{{\bf a)} Graphical representation of the compound $S_A + S_C$ made of two Lattice Proteins. The solid black lines highlight the backbone structures over the lattice cubes. {\bf b)} Theoretical predictions (solid lines) and numerical simulations (dots) on LP data of training and generalization losses as defined in Eq.~\eqref{eq:Losses} (left and right panel, respectively), for two different distributions of negative examples. In practice, we use data $S_A + S_C$ in the bounded state as positives; negatives are data $S_A + S_C$ in the unbounded or mildly bounded state (green and gold, respectively). Here $\alpha^+ = 19$, $\kappa = 5$, $Q = 20$, $L = 54$, with $\vb{M}, \bm{\delta}$ estimated from the data. Numerical simulations are averaged over $20$ trials. }
    \label{fig:app_dimer}
\end{figure*}

Given two structures $S_1$, $S_2$, we collect sequences through MCMC dynamics such that $p_{\text{nat}}(S_1 |\vb{v}_1)$, $p_{\text{nat}}(S_2 |\vb{v}_2) >0.99$. Here we randomly choose two specific compounds, namely $S_A + S_C$ and $S_B + S_C$ (see \cref{fig:app_dimer}a to visualize one of the two compounds). We group lists of sequences to form the dataset used in this work as follows:
\begin{itemize}
    \item sequences $\vb{v}_A$, $\vb{v}_C$ in the bounded state $S_A + S_C$ represent positive data;
    \item sequences $\vb{v}_A$, $\vb{v}_C$ in the unbounded state $S_A + S_C$ represent negative data;
    \item sequences $\vb{v}_B$, $\vb{v}_C$ in the bounded state $S_B + S_C$ or sequences $\vb{v}_A$, $\vb{v}_C$ in the mildly bounded state $S_A + S_C$ represent out-of-sample data.
\end{itemize}

We stress again explicitly that this so-defined model of LPs is not a model with independent features as it involves inter- and intra-structures couplings in the energy terms \eqref{eq:app_En_MJ} and \eqref{eq:app_Int_MJ}. Yet, we find that our theoretical predictions with diagonal covariance matrix closely reproduce numerical simulations (see Fig.~\ref{fig:app_dimer}b).

\section{Additional numerical results}

\subsection{Numerical validation of replica calculation}
\begin{figure*}[t]
    
    \includegraphics[width=\textwidth]{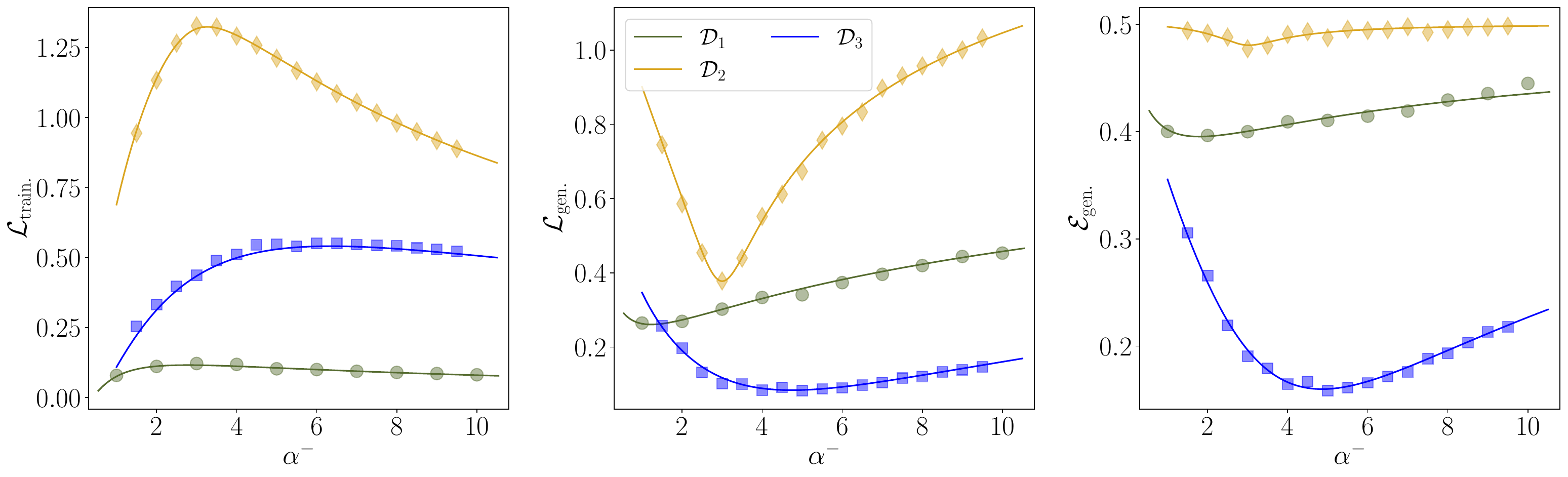}
    \caption{\label{fig:app_rs_theorysims} Theoretical curves with their numerical validation for the quantities $\mathcal{L}_{\text{train.}}$, $\mathcal{L}_{\text{gen.}}$ and $\mathcal{E}_{\text{gen.}}$ -- defined in Eq.~\eqref{eq:Errors},~\eqref{eq:Losses} -- in panels {\bf a)}, {\bf b)} and {\bf c)}, respectively. We test our predictions within three different scenarios, labelled $\mathcal{D}_1$, $\mathcal{D}_2$, $\mathcal{D}_3$ and composed as explained in the text. All numerical experiments are averaged over $20$ trials. }
\end{figure*}

Here we numerically validate our predictions of the quantities $\mathcal{L}_{\text{train.}}$, $\mathcal{L}_{\text{gen.}}$, $\mathcal{E}_{\text{train.}}$ within the RS Ansatz. The numerical solution of the saddle-point equations of interest can be obtained via a simple iterative method and allows to get the theoretical predictions. The code to reproduce the theoretical curves can be found on \href{https://github.com/Eloffredo/RestoringDataBalance}{github}. 
 We check the agreement of numerical simulation with theory over three different synthetic datasets $\mathcal{D}_1$, $\mathcal{D}_2$, $\mathcal{D}_3$, composed as follows. Note that here all the datasets contain diagonal covariance matrices, \ie{} we are setting $C_{i,j}(t,u) = [M_i(t)^2 - M_i(t) M_j(u)] \delta_{ij}$.
\begin{enumerate}[1)]
    \item The dataset $\mathcal{D}_1$ has $L=100$, $Q=2$, $\alpha^+=1$ and $\kappa =0.4$. We sample positive and negative data using $m_i = 0.4$ and $\delta_i = 1$ over each position.
    \item The dataset $\mathcal{D}_2$ has $L=200$, $Q=10$, $\alpha^+ = 3$ and $\kappa = 2$. We sample the tensor $\vb{M}$ as i.i.d. variables uniform in $[0.1,0.3]$ and $\bm{\delta}$ as i.i.d. variables uniform in $[0,0.1]$. We then enforce the normalization conditions setting the last component over each position as 
    \begin{equation}
    M_i(Q) = 1 - \sum_{t=1}^{Q-1} M_i(t), \qquad \delta_i(Q) = - \sum_{t=1}^{Q-1} \delta_i(t).
    \end{equation}
    \item The dataset $\mathcal{D}_3$ has $L=200$, $Q=10$, $\alpha^+ =4.5$ and $\kappa = 2$. We sample the tensors 
    \begin{equation}
    \vb{M}^{\pm}  = \vb{M} \pm \dfrac{\bm{\delta}}{2 \sqrt{L}} \sim \mathcal{N} ( 0.8 \pm 0.2, 0.1),
    \end{equation}
    and rescale them so that on each position the vector sums to one (normalization condition). We then obtain $\vb{M}$, $\bm{\delta}$ from $\vb{M}^{\pm}$ by sum and difference.
\end{enumerate}
We show the results in \cref{fig:app_rs_theorysims}.

\section{On numerical experiments \label{app:num_exps}}
\subsection{Implementation details of RBM}
We train the RBM using a gradient-based method to maximize the log-likelihood $\mathcal{L}_{\text{RBM}} = \sum_{k=1}^P \log p(\vb{v}_k )$, where the sum runs over the training (positive) data. Given a generic parameter $\omega$ to learn, we solve the momentum matching condition
\begin{equation}
    \pdv{\mathcal{L}_{\text{RBM}}}{\omega} = \left< \pdv{E}{\omega}\right>_{\mathrm{M}} - \left< \pdv{E}{\omega}\right>_{\mathrm{D}},
\end{equation}
where the r.h.s. expectation values are over the model and data distribution, respectively. The parameters are updated with persistent-contrastive-divergence (PCD) \citep{tieleman2008training} algorithm. Averages over the model are estimated running $T=100$ parallel Markov chains, while averages over the data are computed on mini-batches of size $T$ sampled from the training set. The RBM weights $w$ are initialized to small random Gaussian values, with a standard deviation equal to $0.1/ \sqrt{L}$, where $L$
is the number of visible units. Eventually, we include a regularization during the training procedure by introducing a L2 penalty term $\mathcal{L}_{\text{RBM}} \to \mathcal{L}_{\text{RBM}} - \lambda \normLtwo{\omega}$. For the MNIST dataset we used the following parameters: $784$ visible units,$100$ hidden units, L2 strenght $0.001$ and $250$ Epochs.

\paragraph{Gibbs Sampling.} To generate a new sample with the RBM, we use Gibbs Sampling as follows
\begin{enumerate}[(i)]
    \item We start from an initial configuration $\vb{v}_0$ (either random or taken from the training set);
    \item We sample the next configuration $\vb{v}_1$ based on $p(\vb{v}_0| \vb{z} )$;
    \item We repeat step (ii) for $K$ times and return the configuration $\vb{v}_K$ as a new sample.
\end{enumerate}
\subsection{Implementation details of supervised classifiers\label{app:classifiers}}

\paragraph{Perceptron.} To validate the theoretical predictions (\eg{} \cref{fig:fig_theory}b), we minimize the following cost function
\begin{equation}
\label{eq:app_cost_perceptron}
    \mathcal{L}(\vb{J},b) = \sum_{\mu} \left( \kappa - y_{\mu} \Delta_{\mu}\right) \theta \left( \kappa - y_{\mu} \Delta_{\mu}\right) - \lambda \sum_{t} \left[ J_i(t) -1 \right]^2 - \lambda \sum_{t} \left( \normLtwo{J} - LQ \right)^2,
\end{equation}
where the last two terms on the r.h.s. enforce the spherical weights and zero-sum condition. We set $\lambda = LQ$ and minimize Eq.~\eqref{eq:app_cost_perceptron} using the L-BFGS-B routine of the Scipy library \citep{jones_scipy:_2001}.
\paragraph{Linear SVM.} To carry out the numerical experiments on MNIST  dataset to compare different mixing balancing strategies (\eg{} \cref{fig:fig_numexps}a) we use Linear Support Vector Machines. We use the Scikit-learn package \citep{pedregosa2011scikitlearn} which implements the SVM as follows
\begin{equation}
    \mathcal{L} (\vb{J},b) = C \sum_{\mu} \max(0,1-y^\mu (\vb{J} \cdot \vb{x}^\mu + b) ) + \dfrac{1}{2} \normLtwo{J},
\end{equation}
where the sum runs over the examples in the dataset. We set the intercept value $b=0$ (option "fit-intercept=False") and the regularization strength $C = 10 $. For each minority class size we average over $100$ realization to make the performance curves smooth enough. We evaluate results using the BA metric. Notice how we can compare numerical curves obtained in this way with theoretical predictions derived from Eq.~\eqref{eq:minimization-SVM}, with $C \to \tilde{\lambda}^{-1}$, $b \to -b$ and rescaling the input data by $\sqrt{L}$.
\paragraph{ResNet-50} We take a pretrained ImageNet ResNet-50 architecture and add a fresh dense layer with $512$ neurons and ReLU activation; we put an output neuron with sigmoid activation for the binary classification task. We fine-tune the network weights on Cifar10 data. We use RMSprop optimizer with a learning rate of $10^{-5}$ and decay of $10^{-5}$ and train for $100$ epochs with a batch-size of $128$.
\subsection{Dataset preprocessing}

\begin{figure*}[h!]
    \centering
    \includegraphics[scale=0.55]{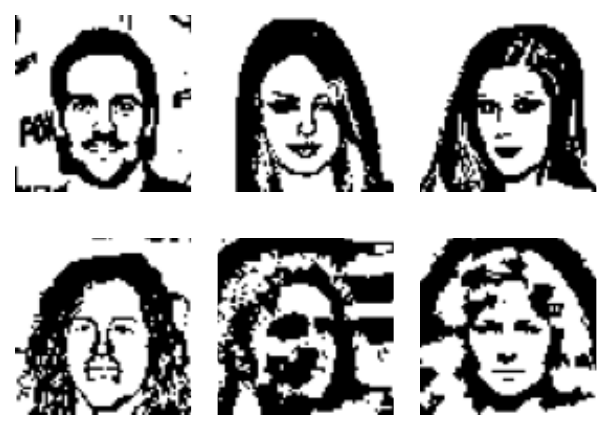}
    \caption{\label{fig:celeba_faces} Randomly selected images of faces from the CelebA dataset after the data preprocessing described in the text. Images in the top row belong to the straight hair class, while bottom row displays faces with wavy hair.}
\end{figure*}

\paragraph{CIFAR10.} CIFAR10 for ResNet-50 fine-tuning are used without any preprocessing. To perform binary classification we split images into two classes according to what they represent -- [airplane, car, bird, cat, deer] and [dog, frog, horse, ship, truck], respectively. The test set is balanced and has size $1000$.
\paragraph{MNIST.} MNIST data are first converted to binary values $\{0,1\}$ setting the threshold value for the black/white mapping at 0.5. To perform binary classification we split the digits in two classes. Parity MNIST (pMNIST) has odd and even digits classes; while 5-MNIST has smaller and larger than digit five classes. Hence, the two classes contain $\sim 30000$ examples each. The test set is balanced and has size $1000$.
\paragraph{FashionMINST.} FMNIST data are first converted to binary values $\{0,1\}$ setting the threshold value for the black/white mapping at 0.5. To perform binary classification we split the dataset in two classes representing "Pullover" and "Shirt" images. Hence, each class has $6000$ examples. The test set is balanced and has size $1000$.
\paragraph{CelebA.} Images of the CelebA datasets are first converted to single (grayscale) channel by taking the average of the three color channels. Then data are binarized through the Sauvola–Pietikäinen adaptive image binarization algorithm \cite{SAUVOLA2000225}, that sets the threshold for the black/white mapping for each pixel $(x,y)$ as 
\begin{equation}
    T(x,y) = m(x,y) \left[ 1 + k \left( \dfrac{s(x,y)}{R} -1 \right) \right].
\end{equation}
Here $m,s$ refer to the mean and standard deviation of the intensity across a window of size $8$ around the pixel $(x,y)$. The two constant values are set as $R = \max s(x,y)$ over all pixels in the image and $k = 0.05$. To perform binary classification we only select images based on two attributes, namely "Straight hair" or "Wavy hair". Some randomly selected faces from the two classes are shown in \cref{fig:celeba_faces}.

\end{document}